\documentclass[aps,pra,twocolumn,superscriptaddress,a4paper,english,longbibliography]{revtex4-1}
\usepackage{silence}
\usepackage{times}
\usepackage{color}
\usepackage[colorlinks=true,urlcolor=blue,citecolor=blue]{hyperref}
\usepackage{url}
\usepackage{soul}
\usepackage{graphicx}
\usepackage{amsmath}
\usepackage{subfigure}
\usepackage{physics}
\usepackage{setspace}
\usepackage{xcolor}
\usepackage{amssymb}
\usepackage{latexsym}
\usepackage{hyperref}% add hypertext capabilities
\DeclareGraphicsExtensions{.png,.eps}
\usepackage{float}
\usepackage{ulem}
\usepackage{appendix}
\usepackage{etoolbox}
\usepackage{fancyhdr}
\usepackage{amsmath}
\usepackage{bm}

\usepackage{xcolor}
\usepackage[urlcolor=blue]{hyperref}      % links to citations
\hypersetup{
    colorlinks = true,                    % text and not border
    citecolor = {blue},
    linkcolor = {purple},
           }
\begin{document}	

\title{Enhancing Quantum Discord in V-shaped Plasmonic Waveguides by Quantum Feedback}

\author{Hua-Wei Zhao}
\affiliation{School of Physics, Beihang University,100191,Beijing, China}

\author{Gen Li}
\affiliation{School of Physics, Beihang University,100191,Beijing, China}
\author{Guo-Feng Zhang}
\email[Corresponding author: ]{gf1978zhang@buaa.edu.cn}
\affiliation{School of Physics, Beihang University,100191,Beijing, China}

\begin{abstract}
	We investigate the impact of a symmetric quantum feedback control on the quantum discord of the $X$ state in V-shaped plasmonic waveguides. Under this feedback, the quantum discord of the Werner state is enhanced from $0$ to $0.38$. This value even continues to rise after reducing the decay rate of the atoms. Furthermore, we get the operational mechanism of feedback control through the evolution of the matrix elements. It confines the initial $4\times 4$ matrix into a $3\times 3$ subspace. As a result, the weights of each ground state in the quantum state change, which suppresses the degradation of Bell state. Lastly, we propose a direction for suggesting an improved feedback Hamiltonian.
\end{abstract}

\maketitle

\section{INTRODUCTION} 

Quantum entanglement plays a crucial role in the development of various quantum technologies. In recent years, significant progress has been achieved in fields such as quantum teleportation~\cite{PhysRevLett.123.070505}, quantum dense coding~\cite{PhysRevApplied.17.064011} and quantum computing~\cite{PhysRevLett.101.040501,PhysRevB.93.125105}. Quantum entanglement has emerged as a valuable resource in these endeavors. However, it is important to note that entanglement is not the only resource to process quantum computation and communication. A notable example is the implementation of Shor's algorithm by IBM Almaden Research Center in 2001~\cite{Vandersypen_2001}. Although entanglement was deemed unsustainable in the experimental environment~\cite{PhysRevLett.83.1054}, the algorithm was successfully executed. In the same year, experiments to perform quantum search without entanglement were also implemented~\cite{PhysRevA.64.042306}. Moreover, it has also been proposed theoretically~\cite{PhysRevLett.83.1054,PhysRevA.72.042316,PhysRevLett.85.2014} and subsequently confirmed through experimentation~\cite{PhysRevLett.101.200501} that certain separable states can outperform their classical counterparts in specific tasks. These findings indicate that entanglement is not the sole quantum correlation that is beneficial for quantum technology. Consequently, it is a natural idea to explore alternative perspectives to characterize and quantify quantum correlations.
    
In classical information theory, the correlation between two variables is described by classical mutual information. This concept is extended to the field of quantum mechanics, the total correlation between two quantum variables or systems contains both classical and quantum correlation. The total correlation is described by the quantum mutual information~\cite{PhysRevA.72.032317}
\begin{equation}\label{QMI}
\begin{split}
T  (\rho^{MN}) = S(\rho^M) + S(\rho^N) - S(\rho^{MN}),
\end{split}
\end{equation}
 where $ \rho^{MN} $ is the total density matrix of the system $ M $ and $ N $, $ \rho^M $ ($ \rho^N $) is the density matrix of system $ M $ ($ N $), and $S(\rho) = -{\rm Tr} (\rho \log _2 \rho)$ is the von Neumann entropy. Quantum mutual information can be expressed as a sum of classical correlation $ C (\rho) $ and quantum correlation $ Q (\rho)$~\cite{PhysRevLett.88.017901,LHenderson_2001,PhysRevA.80.044102,PhysRevA.77.042303}. The quantum part $ Q (\rho) = T (\rho)-C (\rho) $, is defined as quantum discord (QD)~\cite{PhysRevLett.88.017901}. QD, which is not necessarily zero even for some separable states, is a more generalized form of quantum correlation than entanglement.

Like entanglement, QD also gradually dies out due to interaction between the system and environment. Although entanglement and QD are not identical, they both indicate the correlations between two systems, so a natural idea is whether measures to protect quantum entanglement can be used to protect QD. Many ways have been proposed to protect and prolong the entanglement time, such as using feedback control~\cite{PhysRevA.76.010301}, weak measurement~\cite{PhysRevA.95.042342} and transferring quantum states by plasmonic waveguides (PWs)~\cite{Bozhevolnyi2006ChannelPS}.

Surface plasmon photonics is a pivotal branch of nanoscale photonics. PWs break the size limitations of conventional silicon-based optics~\cite{Brongersma2010TheCF}. This feature greatly improves the performance of optical devices. In addition to this, PWs have an interesting property: they enhance the interaction of light with matter~\cite{article}. This attribute has garnered considerable attention in the scientific community. For example, PWs have demonstrated positive effects on prolonging entanglement duration and improving quantum capacity in super dense coding~\cite{PhysRevB.84.235306}. The fabrication of channel waveguides integrated into metal surfaces has been elucidated in Ref.~\cite{Bozhevolnyi2006ChannelPS}. The utilization of surface plasmonic excitations serves the purpose of mitigating the impact of the diffraction limit on electronic and photonic circuits~\cite{PhysRevB.71.195406}. 

On the other hand, several distinct forms of quantum feedback control have been proposed. The homodyne-mediated feedback~\cite{PhysRevA.82.012336,PhysRevA.49.2133} and quantum-jump-based feedback~\cite{PhysRevA.82.012336} controls have been demonstrated to generate steady state entanglement. Continuous measurement feedback has played an active role in non-Markovian entanglement dynamics~\cite{PhysRevA.98.052134}. PT symmetric feedback has been applied in protecting quantum entanglement, QD and quantum fisher information~\cite{Xie2019EnhancingPO}.

This paper introduces two symmetric feedback controls and examines their impact on QD of two-level coupled atomic system in PWs. In addition to analyzing the dynamical evolution of QD, we explore how the parameters of PWs and quantum states influence the stationary QD. Our finding demonstrates that both types of feedback control enhance QD.

This paper is organized as follows: In section II, we explain the theoretical model comprising two coupled qubits mediated by PWs and analysis the QD of Werner state without symmetric feedback control. Section III contains the effect of symmetric feedback control on $X$ states. In section IV, we summarize the similarities and differences between two feedback Hamiltonians and clarify the mechanism behind feedback control. The conclusions are given in section V.

\section{QUANTUM DISCORD WITHOUT FEEDBACK CONTROL}

Consider two identical two-level atoms with spontaneous frequency $ \omega_0 $. Two qubits are situated within an electromagnetic field environment, with a distance $ d $ separating them and a height $ L $ from the bottom of a V-shaped channel waveguide. The metal used to make PWs is silver. The fabrication of such V-shaped PWs has been described in Ref. ~\cite{Bozhevolnyi2006ChannelPS}. The presence of PWs enhances the electromagnetic interaction between the qubits. After applying the Born-Markovian approximation, we can derive the Lindblad equation~\cite{RevModPhys.89.015001}, which makes the dynamical evolution of the system relatively easy to solve. Then by taking trace over environmental degree of freedom, the master equation for the density matrix of the two qubits is written as~\cite{PhysRevA.72.024104}:
\begin{equation}\label{zhufangcheng}
	\begin{split}
		\frac{\partial \rho}{\partial t}=-\frac{i}{\hbar}[H,\rho]+\mathbb{D} [c] \rho ,
	\end{split}
	\end{equation}
where $ \rho $ is the density matrix of the two qubits, the Liouvillian  superoperator $ \mathbb{D} [c] \rho=c \rho c^{\dagger}-c^{\dagger} c \rho/2-\rho c^{\dagger} c/2 $, $ c=-\sqrt{\xi} (\sigma\otimes I + I\otimes \sigma) $ is the jump operator describing the interaction between the systems and environment, $I$ is a $4\times 4$ identity matrix. $ H=\hbar(\omega_0+\delta)( \sigma^{\dagger}\sigma\otimes I+I\otimes \sigma^{\dagger}\sigma) + \hbar\gamma( \sigma^{\dagger}\otimes \sigma + \sigma\otimes \sigma^{\dagger}) $ is the driving Hamiltonian where $\sigma=| g\rangle \langle e| $ and $\sigma^{\dagger}=| e\rangle \langle g| $. $| e\rangle$ is excited state and $| g\rangle$ is ground state of two-level atom. The term $ \xi $ denotes the decay rate of the qubits resulting from their interaction with the environment. The term $ \gamma $ corresponds to  the dipole-dipole coupling induced by PWs. To simplify the calculation, we let $ \omega_0 $ equal to $ \gamma $. The term $ \delta $ is called the Lamb shift, which arises from the interaction between the qubits and the electromagnetic field and can be disregarded when the height $ L $ between the qubits and PWs exceeds $ 10 $ nm~\cite{Hohenester2008InteractionOS}.

\begin{figure}[H]
	\centering
	\subfigure[]{
		\includegraphics[angle=0,width=0.7\linewidth]{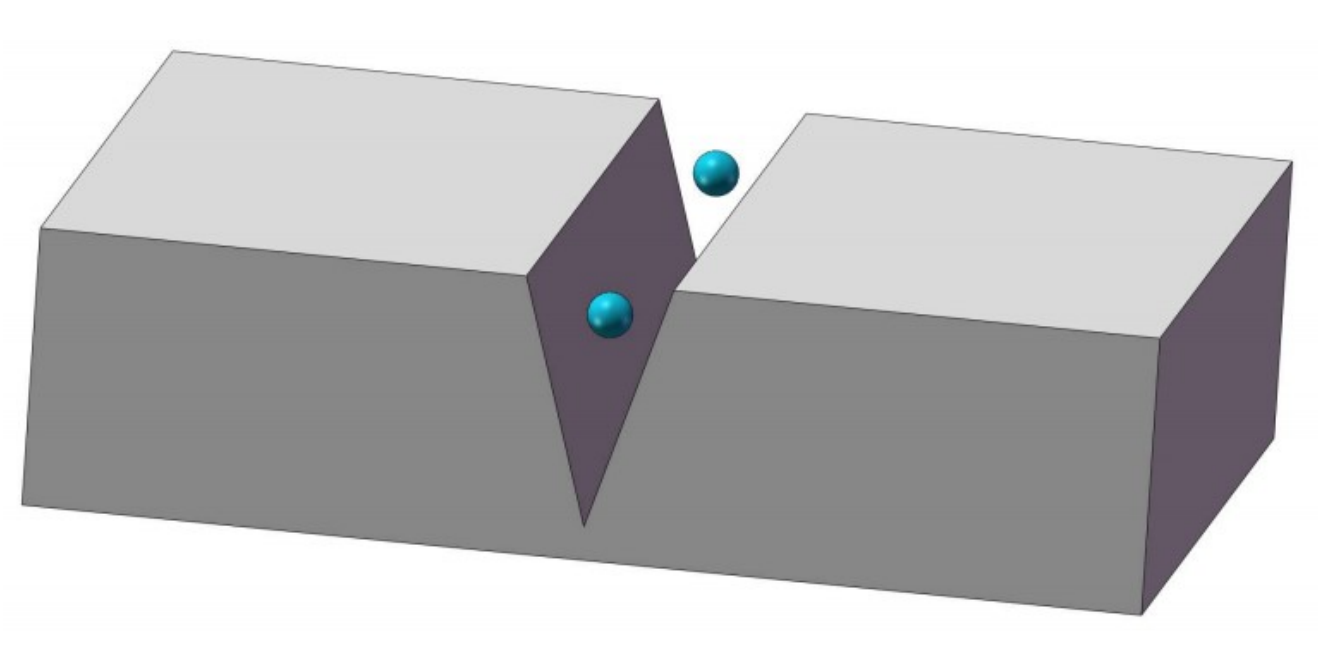}
		\label{subFig1}
	}
	\subfigure[]{
		\includegraphics[angle=0,width=0.7\linewidth]{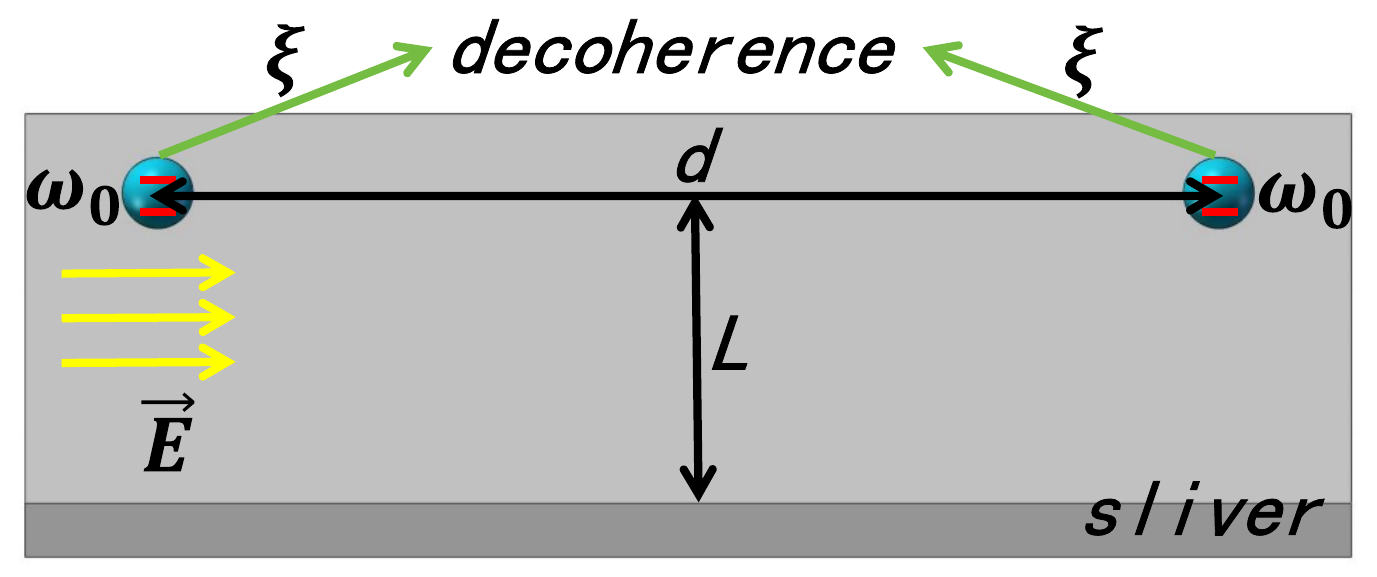}
		\label{subFig2}
	}
	\caption{(Color online) Coupling model of two two-level atoms and electromagnetic field in a silver V-shaped plasmonic waveguide. The tension angle of the V-groove is $40$ degrees. The vertical distance of qubits from the bottom of the groove is $L=150$ nm and the horizontal distance between the two qubits is $d=7L/2=535$ nm. The propagation length is a const $1.7 \mu$m. Under these parameters, $\beta=0.9$. The quantum system will decoherence with the strength $\xi$. (a) 3-dimensional view. (b) side sectional view.}
	\label{Fig7}
\end{figure}

The expressions of $ \xi $ and $ \gamma $ can be obtained by integrating Green's function of the electromagnetic field over a frequency range. They are given as $ \xi\simeq\beta \Xi e^{-d/2l} \cos (k_r d) $ and $ \gamma \simeq \beta \Xi e^{-d/2l} \sin (k_r d)/2 $ in~\cite{MartnCano2010ResonanceET,PhysRevB.82.075427}. The parameters in these two equations can be taken through the plasmonic waveguide. The term $\Xi$ is spontaneous decay rate of atom. Without loss of generality, we let $\Xi=1$. The term $\beta$ also referred as $\beta$-factor gives the plasmonic decay rate as a percentage of the total decay rate. It is one of the most important parameters of PWs, which reflects the effect of PWs on the QD. In Ref. ~\cite{PhysRevB.84.235306}, Moreno et al. proposed a sliver one-dimensional V-shaped plasmonic waveguide. They proved that the optimal value of $\beta$ is 0.9 when $L = 150$ nm and the angle of the V-groove is 40 degrees. Additionally, they calculated that the propagation length of this plasmonic waveguide, denotes as $ l $, has a constant value of $l=(2k_i)^{-1}=1.7$ $\mu$m. $ k_i $ represents the imaginary part of the complex modal wave vector: $k=k_r+ik_i$. Meanwhile, they let the distance between two qubits $d=7L/2$. To simplify calculation, we assume that $ \cos (k_r d)=\sin (k_r d)=\sqrt{2}/2 $. We give this model more graphically in Fig. \ref{Fig7}.

In order to quantify the QD, we need to calculate the total and classical correlations of the system separately. In this section, we focus our discussion on initially prepared arbitrary two-qubit $ X $ states. For $ X $ state we have the following calculation process~\cite{PhysRevA.81.042105}.

In the uncoupled representation ($ |1\rangle =|e,e\rangle$, $|2\rangle =|e,g\rangle$, $|3\rangle =|g,e\rangle$, $|4\rangle =|g,g\rangle $), the density matrix of $ X $ state has the following form:
\begin{equation}\label{Xjuzhen}
	\begin{split}
	\rho^{MN}=
	\begin{pmatrix}
		\rho_{11}&0&0&\rho_{14}\\
		0&\rho_{22}&\rho_{23}&0\\
		0&\rho_{32}&\rho_{33}&0\\
		\rho_{41}&0&0&\rho_{44}\\
	\end{pmatrix},
	\end{split}
\end{equation}
this density matrix can be decomposed with Pauli matrix as
\begin{equation}\label{paolizhen}
	\begin{split}
	&\rho^{MN}=\\
	&\frac{1}{4}[I\otimes  I + \sum\limits_{i=1}\limits^{3}(\alpha_i \sigma^i\otimes \sigma^i )+\alpha_4 I\otimes \sigma^3 + \alpha_5 \sigma^3 \otimes I],
	\end{split}
\end{equation}
where $ \alpha_1=\rho_{14}+\rho_{23}+\rho_{32}+\rho_{41} $, $ \alpha_2=\rho_{23}+\rho_{32}-\rho_{14}-\rho_{41} $, $ \alpha_3=\rho_{11}+\rho_{44}-\rho_{22}-\rho_{33} $, $ \alpha_4=\rho_{11}+\rho_{33}-\rho_{22}-\rho_{44} $ and $ \alpha_5=\rho_{11}+\rho_{22}-\rho_{33}-\rho_{44} $. The four eigenvalues of $ \rho^{MN} $ are
\begin{equation}\label{benzhengzhi}
	\begin{split}
	E_0 =\frac{1}{4}[(1+ \alpha_3 )+\sqrt{(\alpha_4 + \alpha_5)^2 + (\alpha_1 - \alpha_2 )^2}],\\
	E_1 =\frac{1}{4}[(1+ \alpha_3 )-\sqrt{(\alpha_4 + \alpha_5)^2 + (\alpha_1 - \alpha_2 )^2}],\\
	E_2 =\frac{1}{4}[(1- \alpha_3 )+\sqrt{(\alpha_4 - \alpha_5)^2 + (\alpha_1 + \alpha_2 )^2}],\\
	E_3 =\frac{1}{4}[(1- \alpha_3 )-\sqrt{(\alpha_4 - \alpha_5)^2 + (\alpha_1 + \alpha_2 )^2}],\\
	\end{split}
\end{equation}
the total correlation (quantum mutual information) between $M$ and $N$ is given by
\begin{equation}\label{qmi}
	\begin{split}
	T  (\rho^{MN}) = S(\rho^M) + S(\rho^N) + \sum\limits_{i=0}\limits^{3} E_i \log_2 E_i .
	\end{split}
\end{equation}

Ollivier and Zurek proposed to obtain classical correlation by a local von Neumann type measurement~\cite{PhysRevLett.88.017901}. We first take a kind of local measurement $N_k= \{N_0=V | e\rangle \langle e| V^\dagger$, $N_1=V | g\rangle \langle g| V^\dagger \} $ on part $ N $, where $ V=\sum_{i=1}^{3} \epsilon_i \sigma^i $, $ \epsilon_1= \sin(\theta/2)\cos(\phi/2)$, $\epsilon_2= \sin(\theta/2)\sin(\phi/2)$, $\epsilon_3= \cos(\theta/2)$, $0\leqslant \theta \leqslant\pi$, $0\leqslant \phi <2\pi $~\cite{PhysRevA.77.042303}. After local measurement $ {N_k} $, the state of system will change to one of the states
\begin{equation}\label{rho0}
	\begin{split}
	\rho_0 =\frac{1}{2}(I+\sum\limits_{i=1}\limits^{3}q_{0i}\sigma^{i})\otimes (V | e\rangle \langle e| V^\dagger),
	\end{split}
\end{equation}
\begin{equation}\label{rho1}
	\begin{split}
	\rho_1 =\frac{1}{2}(I+\sum\limits_{i=1}\limits^{3}q_{1i}\sigma^{i})\otimes (V | g\rangle \langle g| V^\dagger),
	\end{split}
\end{equation} 
where
\begin{equation}\label{qk1}
	\begin{split}
	q_{k1}= (-1)^k \alpha_1 [\frac{\epsilon_1}{1+(-1)^k \alpha_4 \epsilon_3}],
	\end{split}
\end{equation} 
\begin{equation}\label{qk2}
	\begin{split}
	q_{k2}= (-1)^k \alpha_2 [\frac{\epsilon_2}{1+(-1)^k \alpha_4 \epsilon_3}],
	\end{split}
\end{equation} 
\begin{equation}\label{qk3}
	\begin{split}
	q_{k3}= (-1)^k [\frac{\alpha_3 \epsilon_3+ (-1)^k \alpha_5}{1+(-1)^k \alpha_4 \epsilon_3}].
	\end{split}
\end{equation} 

According to Refs. ~\cite{PhysRevLett.88.017901,PhysRevA.77.042303,PhysRevLett.90.050401}, we can define the classical correlation between parts $ M $ and $ N $ as
\begin{equation}\label{jinddian}
	\begin{split}
	C  (\rho^{MN})=S(\rho^M)-\min\limits_{\{ N_k\} }\{S(\rho^{MN}| {N_k})\},
	\end{split}
\end{equation} 
where $S(\rho^{MN}| {N_k})$ is the quantum conditional entropy of local measurement $N_k$~\cite{PhysRevA.77.042303}. Then, by calculating von Neumann entropy of Eqs. (\ref{rho0}) and (\ref{rho1}), we obtain
\begin{equation}\label{rhokshang}
	\begin{split}
	&S(\rho_k)=\\
	&-\frac{(1+\chi_k)}{2}\log_2 \frac{(1+\chi_k)}{2}-\frac{(1-\chi_k)}{2}\log_2 \frac{(1-\chi_k)}{2},
	\end{split}
\end{equation} 
with
\begin{equation}\label{thetak}
	\begin{split}
	\chi_k= \sqrt{\sum\limits_{i=1}\limits^{3}q^2_{ki}}.
	\end{split}
\end{equation}
Therefore, the classical correlation of $M$ and $N$ can be given by
\begin{equation}\label{cc}
	\begin{split}
	&C  (\rho^{MN})=\\
	&S(\rho^M)-\min\limits_{\{ N_k\} }[\frac{S(\rho_0)+S(\rho_1)}{2}+\alpha_4 \epsilon_3\frac{S(\rho_0)-S(\rho_1)}{2}] .
	\end{split}
\end{equation}
Now we obtain QD as difference between Eqs. (\ref{qmi}) and (\ref{cc}).

Then, we initially prepared two qubits in Werner state which is the degenerate form of Bell state $ |\varPsi\rangle =( |2\rangle+|3\rangle )/\sqrt{2} $. The density matrix is given by:
\begin{equation}\label{WS}
		\begin{split}
		\rho^W(0)&=(1-a)I/4+a|\varPsi\rangle \langle \varPsi |\\
		&=\frac{1}{4}
		\begin{pmatrix}
			1-a&0&0&0\\
			0&1+a&2a&0\\
			0&2a&1+a&0\\
			0&0&0&1-a\\
		\end{pmatrix},
		\end{split}
	\end{equation}
where $ a $ is the probability amplitude of Bell state $ |\varPsi\rangle $.
\begin{figure}[H]
	\centering
	\includegraphics[angle=0,width=0.9\linewidth]{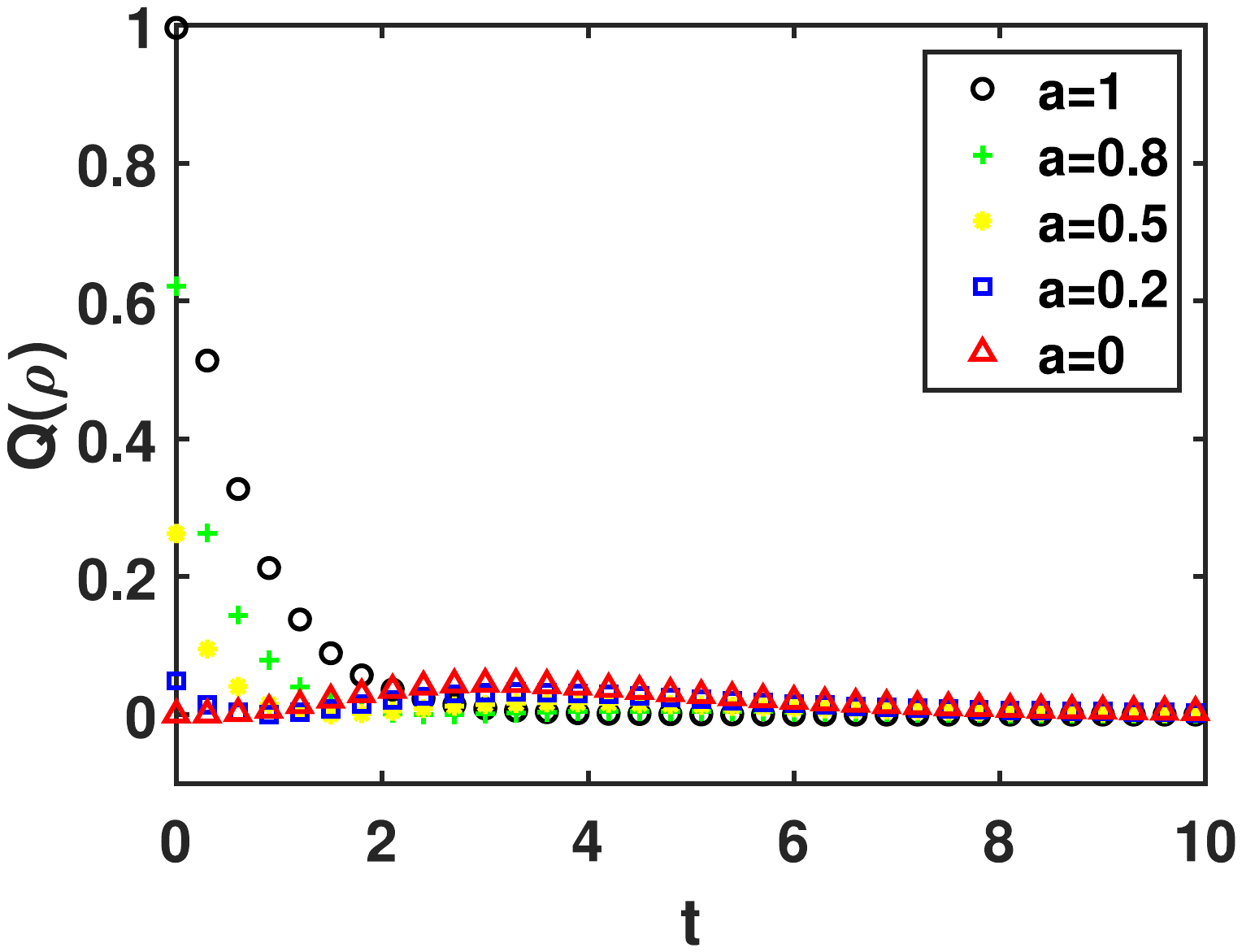}
	\caption{(Color online) Time evolution of QD in V-shaped PWs for initially Werner state. Several crucial parameters take values: $\beta=0.9$, $\xi=0.9\exp (-535/3400)\sqrt{2}/2 $, $\gamma=0.45\exp (-535/3400)\sqrt{2}/2 $, $\omega_0=\gamma$ and $\delta$ has been neglected. $t$ is a dimensionless number.}
	\label{Fig1}
\end{figure}
We can get the dynamical evolution with QD in PWs by using Eqs. (\ref{zhufangcheng}), (\ref{WS}) and the above calculation method in Fig. \ref{Fig1}.

As expected, the interaction between the quantum state and the environment leads to the gradual decay of QD. During our calculations, we observe that the matrix elements are all composed of exponential decay terms. Therefore, the initial $4\times 4$ matrix will gradually converge to $0$ over time. These terms correspond to $\mathbb{D}  [c]\rho$ in Eq. (\ref{zhufangcheng}) which represents the energy dissipation. Although QD vanishes at $t\rightarrow \infty $, we can get valuable information about some special states at $t=0$. When $a=1$, we get the Bell state $|\varPsi\rangle$. Its QD takes maximum value $1$. When $a=0$, the Werner state becomes a mixed state $(|1\rangle \langle1|+|2\rangle \langle2|+|3\rangle \langle3|+|4\rangle \langle4|)/4$. Its density matrix is a diagonal matrix:
\begin{equation}\label{mixed state}
	\begin{split}
	\rho^W(t=0,a=0)
	=\frac{1}{4}
	\begin{pmatrix}
		1&0&0&0\\
		0&1&0&0\\
		0&0&1&0\\
		0&0&0&1\\
	\end{pmatrix}.
	\end{split}
\end{equation}
The QD of this mixed state is $0$. We also easily verify that four pure states $|1\rangle$, $|2\rangle$, $|3\rangle$ and $|4\rangle$ all have no quantum correlation. When the main diagonal elements appear alone, they do not contribute to QD. This observation conveys a crucial message to us: $\rho_{23}$ and $\rho_{32}$ are key points for QD.

\section{QUANTUM DISCORD UNDER SYMMETRIC QUANTUM FEEDBACK CONTROL}

In this section, we explore the application of quantum feedback control to the quantum system. Experimentally, there are two main problems involved in realizing feedback control: under which conditions to apply feedback and how to apply the feedback operation to qubits.

For the first problem, a homodyne detection device that relies on photodetection is one of the most efficient experimental scheme~\cite{PhysRevA.49.2133,PhysRevA.43.3832,PhysRevLett.125.060404}. As we have mentioned in the previous two sections, QD is caused by energy dissipation due to the decoherence of qubits. Physically, the energy dissipation manifests itself as the release of photons from atoms to the environment. We connect a cavity to the plasmonic waveguide as an output field for photons. The homodyne detector is coupled to the output field and performs periodic measurement to the field. When the frequency of the measurement is high enough, we can approximate that the detector is monitoring the decoherence at all time~\cite{PhysRevA.49.2133}. Once decoherence occurs in qubits, the photons are released from the atoms into the output cavity and are subsequently detected by the homodyne detector. Upon detecting the photons, the homodyne detector responds with a current signal $ I $. This current signal is the precondition for the feedback operation $ F $. The term $F$ is the operator of the feedback control applied to the quantum system.

In Ref. ~\cite{PhysRevA.49.2133,PhysRevLett.70.548,PhysRevA.47.642}, H. M. Wiseman and G. J. Milburn studied the evolution of qubits after the homodyne measurement. Subsequently, the master equation of the quantum system and the environment was obtained from the output of the plasmonic waveguide. Finally, after applying the feedback operator $F$ and taking partial trace to the degree of freedom of the environment, the feedback master equation of quantum system can be obtained:~\cite{PhysRevA.77.012339,PhysRevA.78.012334}
\begin{equation}\label{fankuizhufangcheng}
	\begin{split}
		\frac{\partial\rho}{\partial t}=-\frac{i}{\hbar}[H+\frac{1}{2} (c^{\dagger}F+Fc) ,\rho]+\mathbb{D} [c-iF] \rho.
	\end{split}
\end{equation}
The $ F $ is the feedback control operator that we impose on the qubits. We propose a feedback operator that has symmetry for two atoms~\cite{PhysRevA.77.012339}: 
\begin{equation}\label{fankui}
	\begin{split}
		\bm{F = \mu(\sigma_x\otimes \sigma_z + \sigma_z\otimes \sigma_x)-(\sigma_x\otimes I + I\otimes \sigma_x)}.
	\end{split}
\end{equation}

The feedback control affects the master equation in two ways: by replacing $c$ with $c-iF$, and by adding an extra term $(c^{\dagger}F+Fc)/2$ to the driving Hamiltonian. 

\begin{figure}[H]
	\centering
	\includegraphics[angle=0,width=0.9\linewidth]{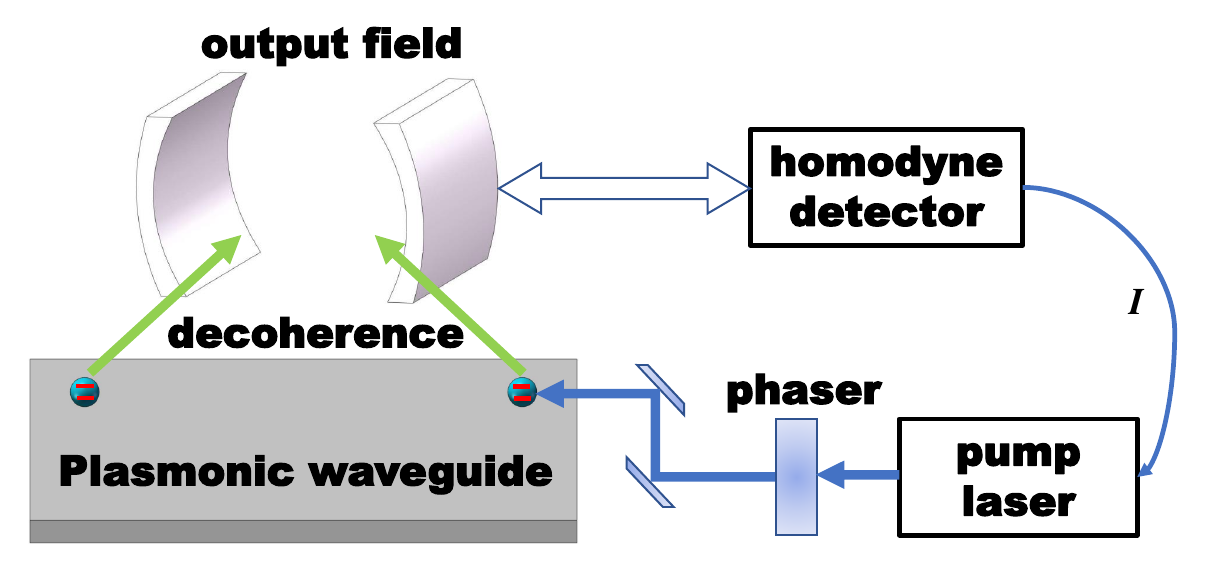}
	\caption{(Color online) Diagram of experimental apparatus to realize feedback control by homodyne detection and pump laser. We merely plot the control of a laser over a single qubit. In fact, after obtaining the current signal $I$ from the homodyne detector, it is easy to control two pump lasers at the same time through computer and circuit.}
	\label{Fignew}
\end{figure}

Experimentally, $\sigma_x$ and $\sigma_z$ represent quantum $X$ gate and quantum $Z$ gate. To realize the quantum operation $F$, selecting a suitable physical system to build quantum logic gates is the key point. The essence of quantum logic gates is to make the quantum system occur unitary evolution toward specified directions. The quantum gate systems that have been proposed include the ion trap scheme~\cite{Steane1996TheIT,PhysRevLett.75.4714,PhysRevLett.74.4091}, the cavity quantum electrodynamics scheme~\cite{PhysRevLett.75.4710,PhysRevLett.74.4087}, the quantum dot scheme~\cite{PhysRevLett.74.4083,PhysRevA.57.120} and laser scheme~\cite{Knill2001ASF,Marshall09}. For qubits based on energy level or spin, we can utilize the interaction between single-mode laser field and atoms to realize quantum logic gate operations. We only need to adjust the amplitude and frequency of the single-mode laser to realize the arbitrary rotation operation of the qubits. In addition, through phaser or the interaction between a short-pulse laser and the second-order virtual photon term of an atom, we can impose a tunable phase on the qubits~\cite{PhysRevA.2.883,PhysRevA.2.889,PhysRevA.52.636}. By this approach we can easily realize $\sigma_x$ and $\sigma_z$. Therefore, we only need the homodyne detector to be linked to the pump laser. Once quantum decoherence occurs, the homodyne current can control the pump laser to apply feedback operator $F$ on the quantum system. In Fig. \ref{Fignew}, we give a visual presentation of the above experimental process.

In what follows, we study the effect of $F$ on QD. The term $ \mu $ takes values $\mu\in [-1,0)\cup (0,1]$. When $\mu=0$, it reduces to a feedback Hamiltonian that has been explored in Ref. ~\cite{PhysRevA.71.042309}. So we consider two extreme cases, $ \mu = -1 $ and $ \mu = 1 $, respectively.

\subsection{Feedback Control $F_1$: $\mu = -1$}

When $ \mu = -1 $, the feedback Hamiltonian is~\cite{PhysRevA.72.024104}
\begin{equation}\label{-1fankui}
	\begin{split}
		F_1 = -\sigma_x\otimes \sigma_z - \sigma_z\otimes \sigma_x-\sigma_x\otimes I - I\otimes \sigma_x .
	\end{split}
\end{equation}

\begin{figure}[H]
	\centering
	\includegraphics[angle=0,width=0.9\linewidth]{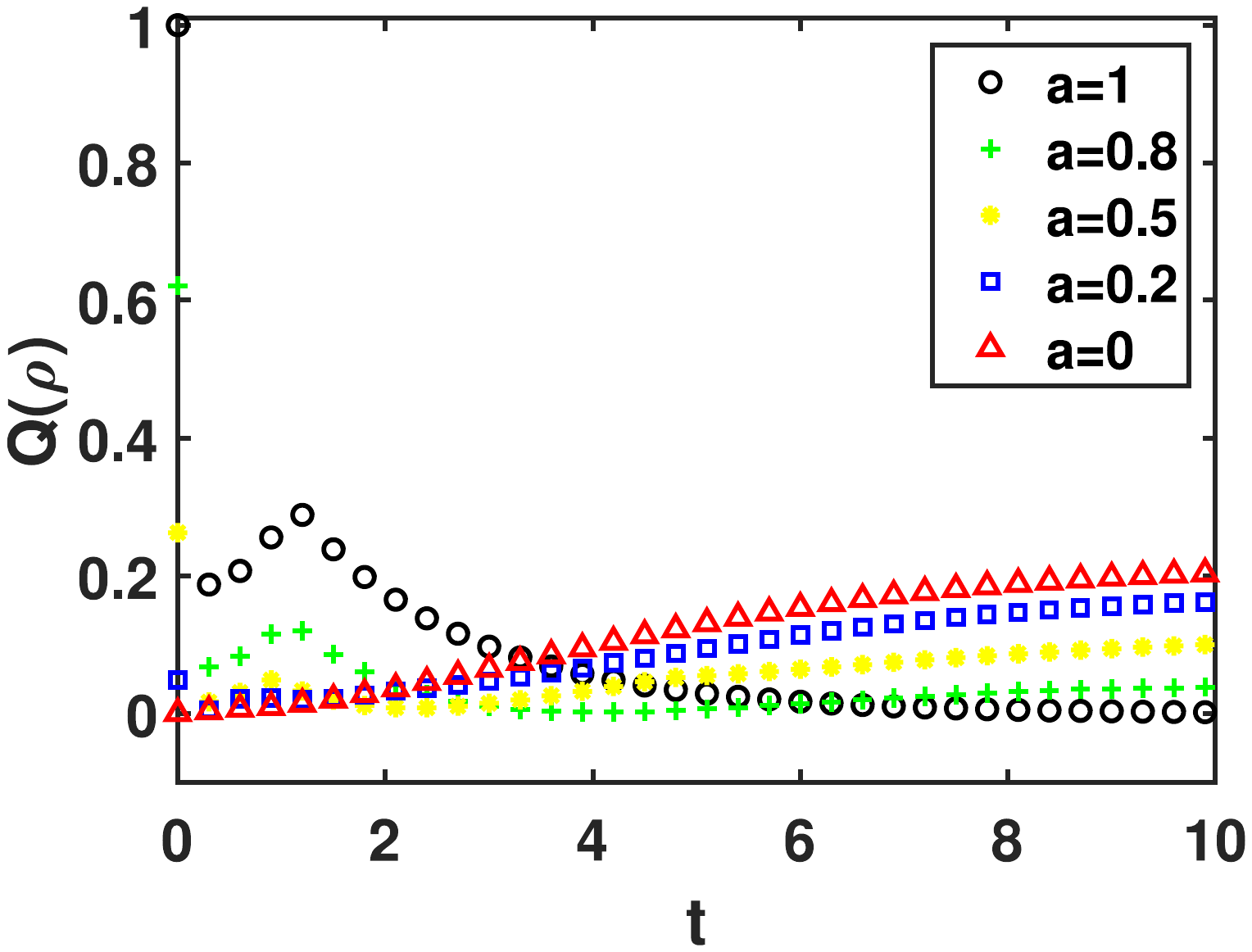}
	\caption{(Color online) Time evolution of QD in V-shaped PWs for initially Werner state under symmetric quantum feedback $ F_1 $. Several crucial parameters take values: $\mu=-1$, $\beta=0.9$, $\xi=0.9\exp (-535/3400)\sqrt{2}/2 $, $\gamma=0.45\exp (-535/3400)\sqrt{2}/2 $, $\omega_0=\gamma$ and $\delta$ has been neglected. $t$ is a dimensionless number.}
	\label{Fig2}
\end{figure}

Through Eqs. (\ref{fankuizhufangcheng}), (\ref{WS}) and (\ref{-1fankui}), we can obtain the evolution of the matrix elements of the Werner State under this feedback control:

\begin{flalign}\label{f1rho11}
	\begin{split}
		& \rho^W_{11} (t)_{F_1}=\\
		& \frac{\mathcal{C} _1-\mathcal{D} _1}{4(4+\xi)} [(1-a)(4+\xi)\mathcal{A} _1+2(1+3a)\sqrt{4+\xi} \mathcal{B} _1] ,
	\end{split}&
\end{flalign}

\begin{flalign}\label{f1rho22}
	\begin{split}
		&\rho^W_{22} (t)_{F_1}=\rho^W_{33} (t)_{F_1}=\frac{\mathcal{C} _1-\mathcal{D} _1}{16} [2(1+3a)\mathcal{A} _1 \\
		&+(1-a)\sqrt{4+\xi} \mathcal{B} _1+2(1-a)(\mathcal{C} _1+\mathcal{D} _1)] ,
	\end{split}&
\end{flalign}

\begin{flalign}\label{f1rho23}
	\begin{split}
		&\rho^W_{23}(t)_{F_1}=\rho^W_{32}(t)^{\ast }_{F_1}=\\
		&\frac{(1+i)(\mathcal{C} _1-\mathcal{D} _1)}{32} [(2-2i)(1+3a)\mathcal{A} _1\\
		&+(1-a)\sqrt{-2i(4+\xi)} \mathcal{B} _1-(1-a)(2-2i)(\mathcal{C} _1+\mathcal{D} _1)] ,
	\end{split}&
\end{flalign}

\begin{flalign}\label{f1rho44}
	\begin{split}
		&\rho^W_{44}(t)_{F_1}=-\frac{(1+i)(\mathcal{C} _1-\mathcal{D} _1)}{16(4+\xi)} [(4-4i)(1+a)(4+\xi)\mathcal{G} _1\\
		&+(8+8a-a\xi+\xi)\sqrt{-2i(4+\xi)} \mathcal{J} _1] ,
	\end{split}&
\end{flalign}
where $ \mathcal{A} _1=\cosh (4t\sqrt{4+\xi}) $, $ \mathcal{B} _1=\sinh (4t\sqrt{4+\xi}) $, $ \mathcal{C} _1=\cosh [2t(4+\xi)] $, $ \mathcal{D} _1=\sinh [2t(4+\xi)] $, $ \mathcal{G} _1=\cosh [(2+2i)t\sqrt{-2i(4+\xi)}] $, $ \mathcal{J} _1=\sinh [(2+2i)t\sqrt{-2i(4+\xi)}] $. The expression for $\rho^W_{14}(t)_{F_1}$ and $\rho^W_{41}(t)_{F_1}$ are not listed here due to their verboseness. The rest matrix elements remain $0$, which indicates that the feedback control does not change the shape of the matrix during the evolution. Through the calculation method of QD in section II, we can obtain the time evolution of QD for the Werner State in Fig. \ref{Fig2}.

By comparing Figs. \ref{Fig1} and \ref{Fig2}, we observe that QD does not gradually converge to $0$ under the feedback control $F_1$. The curve exhibits significant fluctuations when $t<4$ s. However, the QD gradually stabilizes at $t=10$ s. This suggests that this particular feedback control has a positive impact on preserving QD. However, the degree of protection varies for different value of parameter $a$. It can be speculated from Fig. \ref{Fig2} that the protective effect of $F_1$ on the stationary QD of Werner state is inversely proportional to the value of $a$. In order to better analyze this effect, we make $t\rightarrow \infty$ in Eqs. (\ref{f1rho11}), (\ref{f1rho22}), (\ref{f1rho23}) and (\ref{f1rho44}) to obtain the stationary matrix of the Werner State under $F_1$:
\begin{equation}\label{WSF1wentai}
	\begin{split}
	\rho^W(t\rightarrow \infty)_{F_1}
	=\frac{1}{8}
	\begin{pmatrix}
		0&0&0&0\\
		0&1-a&-1+a&0\\
		0&-1+a&1-a&0\\
		0&0&0&6+2a\\
	\end{pmatrix}.
	\end{split}
\end{equation}

\begin{figure}[H]
	\centering
	\includegraphics[angle=0,width=0.9\linewidth]{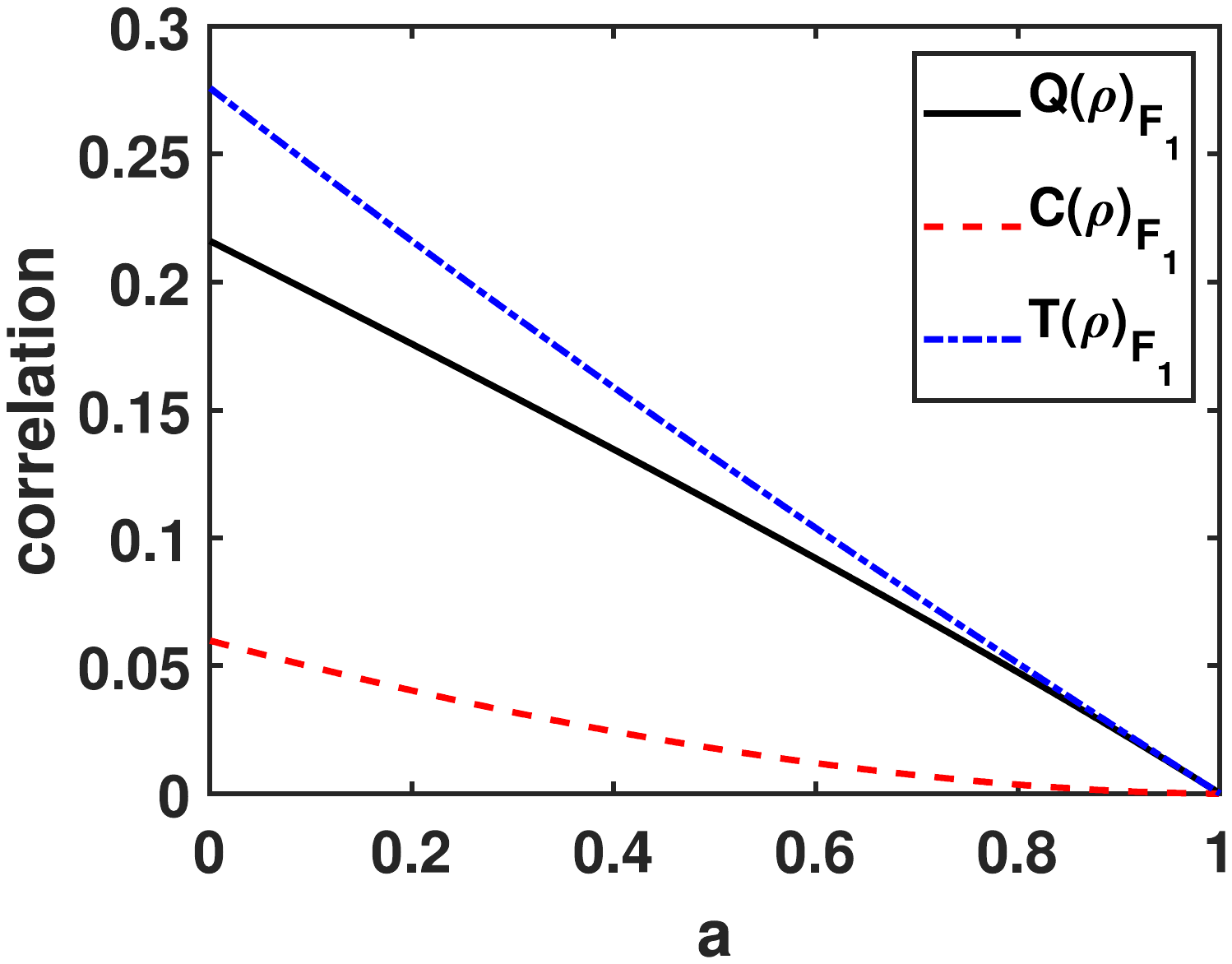}
	\caption{(Color online) Curve of stable total correlation $T (\rho)_{F_1}$ (blue dashdotted line), classical correlation $C (\rho)_{F_1}$ (red dashed line) and QD $Q (\rho)_{F_1}$ (black solid line) about initial parameter $a$ of Werner state under $F_1$. }
	\label{Fig3}
\end{figure}

Now we can obtain stable total correlation and classical correlation through Eqs. (\ref{qmi}), (\ref{cc}) and (\ref{WSF1wentai}):
\begin{flalign}\label{Iguanlian1}
	\begin{split}
		T (\rho)_{F_1}=4-\frac{7+a}{4}\log _2 (7+a)+\frac{3+a}{4}\log _2 (3+a),
	\end{split}&
\end{flalign}

\begin{flalign}\label{Cguanlian1}
	\begin{split}
		&C  (\rho)_{F_1}=-1-\frac{1-a}{8}\log _2 (1-a)-\frac{7+a}{8}\log _2 (7+a)\\
		&+\frac{2+\sqrt{2\Delta }}{16} \log _2 (8+\sqrt{2\Delta })+\frac{2-\sqrt{2\Delta }}{16} \log _2 (8-\sqrt{2\Delta }),
	\end{split}&
\end{flalign}
where $\Delta=4a^2+8a+20$. To separately know the dependence of three kinds of correlations on $a$, we plot Fig. \ref{Fig3}.

Based on the comparison between Fig. \ref{Fig2} and \ref{Fig3}, we conclude that $F_1$ can enhance the QD of Werner state when $a\neq 1$ and the effect is inversely proportional to the value of $a$. The inverse proportionality arises from the fact that both $T (\rho)_{F_1}$ and $C  (\rho)_{F_1}$ decrease monotonically with increasing $a$. More importantly, the decay rate of $T (\rho)_{F_1}$ is significantly higher than that of $C (\rho)_{F_1}$. Now we want to know that how does $F_1$ enhance QD and why does QD become $0$ when $a=1$.

 We find that the original matrix is localized into a $3\times 3$ subspace after converging to the steady state. The negative exponential term $\mathcal{C} _1-\mathcal{D} _1$ are common coefficients of $\rho^W_{11} (t)_{F_1}$, $\rho^W_{22} (t)_{F_1}$, $\rho^W_{33} (t)_{F_1}$, $\rho^W_{44} (t)_{F_1}$, $\rho^W_{23} (t)_{F_1}$ and $\rho^W_{32} (t)_{F_1}$. It is easy to prove that the terms containing $\mathcal{C} _1+\mathcal{D} _1$, $\mathcal{G} _1$ and $\mathcal{J} _1$ happen to generate constant terms when multiplied by the $\mathcal{C} _1-\mathcal{D} _1$. And the terms containing $\mathcal{A} _1$ and $ \mathcal{B} _1$  remain negative exponential terms. The final stable matrix is made up of these constant terms. In Section II we have pointed out that the matrix element without feedback consists entirely of the negative exponential terms of $e$. The use of feedback preserves some constant terms in the matrix, which is why $F_1$ enhances QD. We have verified that for pure states $| 1\rangle$ or $| 4\rangle$, their QD $=0$. When $a=1$, the steady state becomes a pure state $| 4\rangle \langle 4| $. This explains why QD eventually becomes $0$ in this case.

Besides, we replace the Werner state with an arbitrary $X$ state:
\begin{equation}\label{Xstate}
	\begin{split}
	\rho^X(0)_{F_1}=
	\begin{pmatrix}
		\rho_{11}(0)&0&0&\rho_{14}(0)\\
		0&\rho_{22}(0)&\rho_{23}(0)&0\\
		0&\rho_{32}(0)&\rho_{33}(0)&0\\
		\rho_{41}(0)&0&0&\rho_{44}(0)\\
	\end{pmatrix}.
	\end{split}
\end{equation}
After the same process, we obtain its stationary matrix:
\begin{equation}\label{-1Xstatewentai}
	\begin{split}
	&\rho^X(t\rightarrow \infty)_{F_1}\\
	&=\frac{1}{4}
	\begin{pmatrix}
		0&0&0&0\\
		0&\kappa_{1}-\kappa_{2}&\kappa_{2}-\kappa_{1}&0\\
		0&\kappa_{2}-\kappa_{1}&\kappa_{1}-\kappa_{2}&0\\
		0&0&0&2\kappa_{1}+2\kappa_{2}+4\kappa_{3}\\
	\end{pmatrix},
	\end{split}
\end{equation}
where $\kappa_{1}=\rho_{22}(0)+\rho_{33}(0)$, $\kappa_{2}=\rho_{23}(0)+\rho_{32}(0)$ and $\kappa_{3}=\rho_{11}(0)+\rho_{44}(0)$. At this point, we can draw a more general conclusion: when an $X$ state satisfies the condition $ \rho_{22}(0)+\rho_{33}(0)-\rho_{23}(0)-\rho_{32}(0)\neq 0 $, the feedback control $F_1$ will have protective effect on QD.

\subsection{Feedback Control $F_2$: $ \mu = 1 $}

When $ \mu = 1 $, the feedback Hamiltonian is
\begin{equation}\label{1fankui}
	\begin{split}
		F_2 = \sigma_x\otimes \sigma_z + \sigma_z\otimes \sigma_x-\sigma_x\otimes I - I\otimes \sigma_x .
	\end{split}
\end{equation}
Through the same process, we obtain expressions for the evolution of the matrix elements of the Werner state under $F_2$:
\begin{flalign}\label{f2rho11}
	\begin{split}
		\rho^W_{11} (t)_{F_2}=\frac{1}{4} (1-a) e^{-2t\xi} ,
	\end{split}&
\end{flalign}

\begin{flalign}\label{f2rho22}
	\begin{split}
		&\rho^W_{22} (t)_{F_2}=\rho^W_{33} (t)_{F_2}=\\
		&\frac{1}{64}[\frac{8(20+\xi-4a-a\xi)}{8+\xi}+\mathcal{A} _2+\mathcal{B} _2-\mathcal{C} _2+\mathcal{D} _2],
	\end{split}&
\end{flalign}

\begin{flalign}\label{f2rho23}
	\begin{split}
		&\rho^W_{23}(t)_{F_2}=\rho^W_{32}(t)^{\ast }_{F_2}=\\
		&\frac{1}{64}[8(-1+a+\frac{12+4a}{8+\xi})+\mathcal{A} _2+\mathcal{B} _2-\mathcal{C} _2+\mathcal{D} _2],
	\end{split}&
\end{flalign}

\begin{flalign}\label{f2rho44}
	\begin{split}
		&\rho^W_{44}(t)_{F_2}=\frac{1}{32}[\frac{8(3+a)(4+\xi)}{8+\xi}-\mathcal{A} _2-\mathcal{B} _2+\mathcal{G} _2-\mathcal{D} _2],
	\end{split}&
\end{flalign}

\begin{flalign}\label{f2rho14}
	\begin{split}
		&\rho^W_{14}(t)_{F_2}=\rho^W_{41}(t)^{\ast }_{F_2}=\\
		&\frac{(1-i)(1-a)\sqrt{\xi}(e^{-2t\xi}-e^{-t(4+(1+i)\xi)})}{-2-2i+\xi},
	\end{split}&
\end{flalign}
where $\mathcal{A} _2=64\xi i(-1+a)e^{-t[4+(1+i)\xi]}/[(-2-2i+\xi)(6+6i+\xi)]$, $\mathcal{B} _2=64\xi(-1+a)e^{(-1+i)(2+2i+\xi)t}/[-24+(4+4i)\xi+i\xi^2]$, $\mathcal{C} _2=(-1+a)e^{-2t\xi}(-32-8\xi+\xi^3)/[8+(-4+\xi)\xi]$, $\mathcal{D} _2=e^{-2t(8+\xi)}[-\xi(480+\xi(184+\xi(16+\xi)))+a(9216+\xi(4320+\xi(696+\xi(48+\xi))))]/[(6-6i+\xi)(6+6i+\xi)(8+\xi)]$, $\mathcal{G} _2=(-1+a)e^{-2t\xi}(32-40\xi+8\xi^2+\xi^3)/[8+(-4+\xi)\xi]$. The other matrix elements remain $0$. This means that $F_2$ also does not change the shape of the matrix. The matrix elements $\rho^W_{22} (t)_{F_2},\rho^W_{33} (t)_{F_2},\rho^W_{23} (t)_{F_2},\rho^W_{32} (t)_{F_2}$ and $\rho^W_{44} (t)_{F_2}$ share a common characteristic: they all consist of a constant term and several exponential decay terms. These exponential terms decay to $0$ over time, reflecting the energy dissipation of the quantum state as it interacts with environment. The constant term represents the impact of feedback control on the quantum state. Based on the characteristics of the matrix elements, we can anticipate that feedback control $F_2$ is effective in maintaining QD. So we draw the evolution curve of QD in the presence of $F_2$.
\begin{figure}[H]
	\centering
	\includegraphics[angle=0,width=0.9\linewidth]{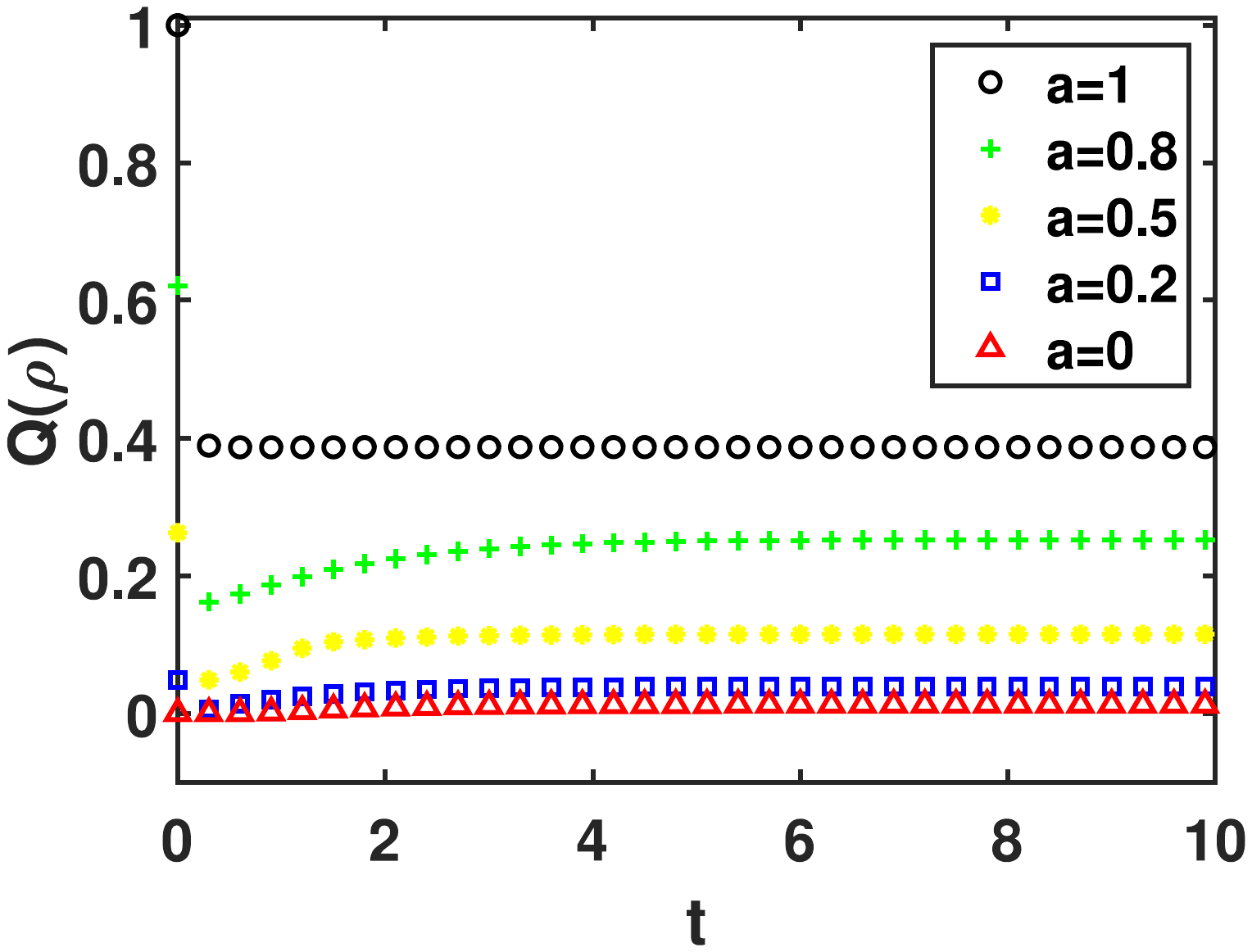}
	\caption{(Color online) Time evolution of QD in V-shaped PWs for initially Werner state under symmetric quantum feedback $ F_2 $. Several crucial parameters take values: $\mu=1$, $\beta=0.9$, $\xi=0.9\exp (-535/3400)\sqrt{2}/2 $, $\gamma=0.45\exp (-535/3400)\sqrt{2}/2 $, $\omega_0=\gamma$ and $\delta$ has been neglected. $t$ is a dimensionless number.}
	\label{Fig4}
\end{figure}

In Fig. \ref{Fig4}, it can be observed that  $F_2$ exhibits a more pronounced effect on enhancing QD. The quantum correlation under $F_2$ does not show superfluous fluctuations and rapidly reaches a steady value at $t = 0.1$ s. When $a = 1$, this steady value reaches $0.38$. The maximum enhancement effect of $F_2$ is twice as large as that of $F_1$. However, the impact of parameter $a$ on the protection effect seems to be completely opposite under feedback control $F_1$ and $F_2$. For this reason we again let $t\rightarrow \infty $, and derive the stationary matrix of the Werner state:
\begin{equation}\label{WSF2wentai}
	\begin{split}
	&\rho^X(t\rightarrow \infty)_{F_2}=\\
	&\begin{pmatrix}
		0&0&0&0\\
		0&\frac{20+\xi-a(4+\xi)}{8(8+\xi)}&\frac{4-\xi+a(12+\xi)}{8(8+\xi)}&0\\
		0&\frac{4-\xi+a(12+\xi)}{8(8+\xi)}&\frac{20+\xi-a(4+\xi)}{8(8+\xi)}&0\\
		0&0&0&\frac{(3+a)(4+\xi)}{4(8+\xi)}\\
	\end{pmatrix}.
	\end{split}
\end{equation}

Unsurprisingly, the Werner state is localized to a $3\times 3$ subspace again. The difference is that these matrix elements contain not only the parameter $a$, but also decay rate $\xi$. Then we can give the expressions of correlations:
\begin{flalign}\label{Iguanlian2}
	\begin{split}
		&T (\rho)_{F_2}=-\frac{1}{8\Delta_1}[2\Delta_2\log_2\frac{\Delta _2}{8\Delta _1}-\Delta_3\log_2\frac{\Delta _3}{4\Delta _1}\\
		&+2\Delta_4\log_2\frac{\Delta _4}{8\Delta _1}-8(3+a)\log_2\frac{3+a}{\Delta _1}\\
		&-2(1-a)\Delta _1 \log_2\frac{1-a}{4}],
	\end{split}&
\end{flalign}
\begin{flalign}\label{Cguanlian2}
	\begin{split}
		&C (\rho)_{F_2}=-\frac{1}{8\Delta_1}[\Delta_2\log_2\frac{\Delta _2}{8\Delta _1}+\Delta_4\log_2\frac{\Delta _4}{8\Delta _1}]\\
		&+\frac{1}{16\Delta_1}[(\Delta_4+\Delta_5)\log_2(\Delta_4+\Delta_5)-2\Delta _4\log_2 \Delta _4\\
		&+(\Delta_4-\Delta_5)\log_2(\Delta_4-\Delta_5)-2\Delta _4 ],
	\end{split}&
\end{flalign}
where $\Delta _1=8+\xi$, $\Delta _2=20+\xi-a(4+\xi)$, $\Delta _3=(3+a)(4+\xi)$, $\Delta _4=44+7\xi+a(4+\xi)$ and $\Delta _5=4+5\xi+3a(4+\xi)$. In the case $\xi=0.9e^{-525/3400}\sqrt{2}/2$, we give the curve of three correlations with respect to $a$.
\begin{figure}[H]
	\centering
	\includegraphics[angle=0,width=0.9\linewidth]{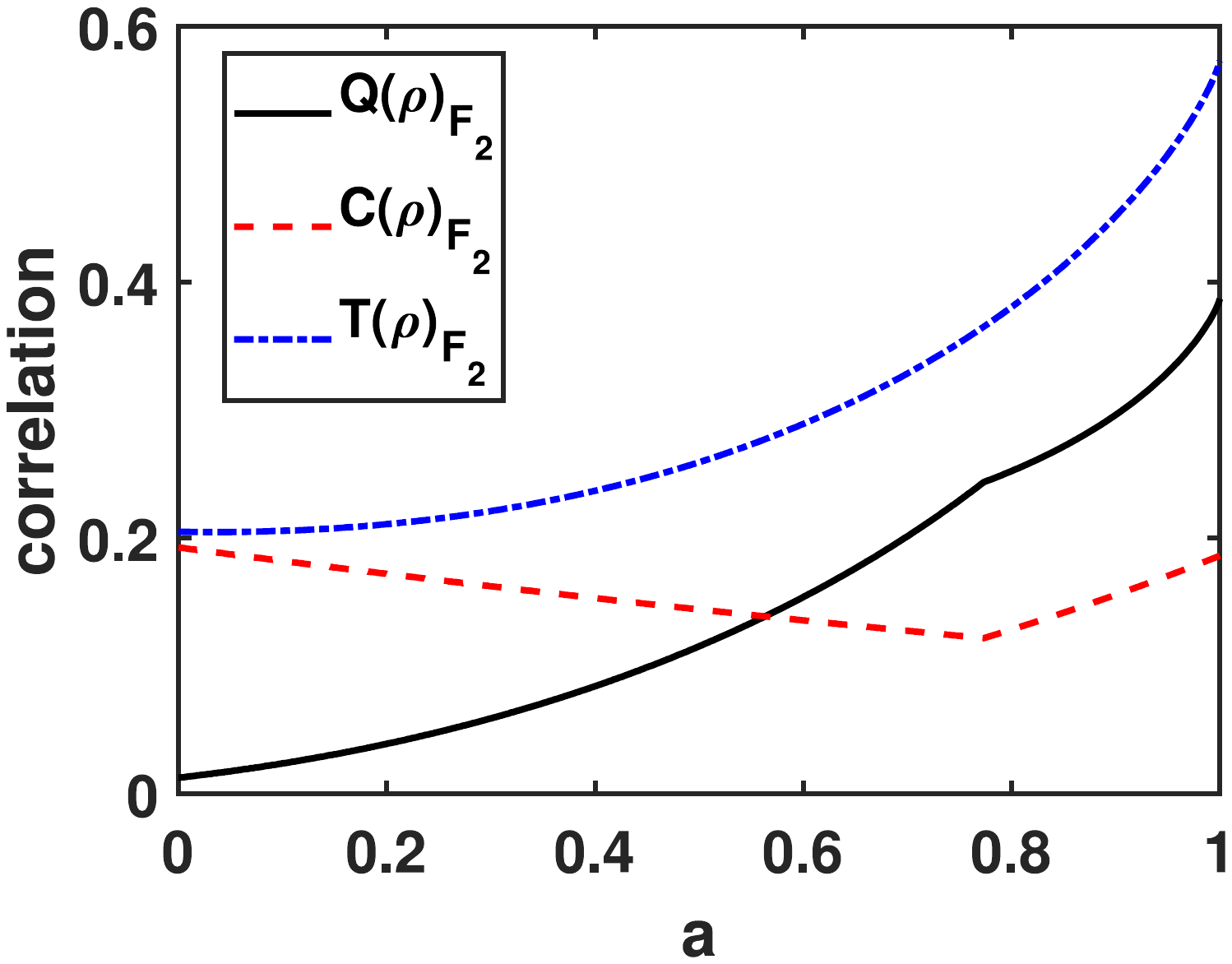}
	\caption{(Color online) Curve of stable total correlation $T (\rho)_{F_2}$ (blue dashdotted line), classical correlation $C (\rho)_{F_2}$ (red dashed line) and quantum discord $Q (\rho)_{F_2}$ (black solid line) about initial parameter $a$ of Werner state under $F_2$.}
	\label{Fig5}
\end{figure}

Fig. \ref{Fig5} illustrates that for a given parameter $a$, the protective effect of $F_2$ on QD of Werner state is proportional to $a$. As $a$ increases, $T(\rho)_{F_2}$  also increases monotonically. However, $C(\rho)_{F_2} $ shows a decreasing and then increasing trend. This turning point occurs at $a = 0.77$. Fortunately, the growth rate of $T(\rho)_{F_2} $ increases rapidly after the turning point. So QD continues to grow steadily, although it is affected at $a=0.77$. This is a completely opposite result to $F_1$. This opposability comes from the two extreme values of $\mu$. On the one hand, the QD reaches a maximum value $0.38$ when $a=1$. On the other hand, the QD does not become $0$ when $a=0$. Thus, we think $F_2$ is a superior feedback to $F_1$ in enhancing QD of Werner state in Eq. (\ref{WS}) overall.

\begin{figure}[H]
	\centering
	\includegraphics[angle=0,width=1\linewidth]{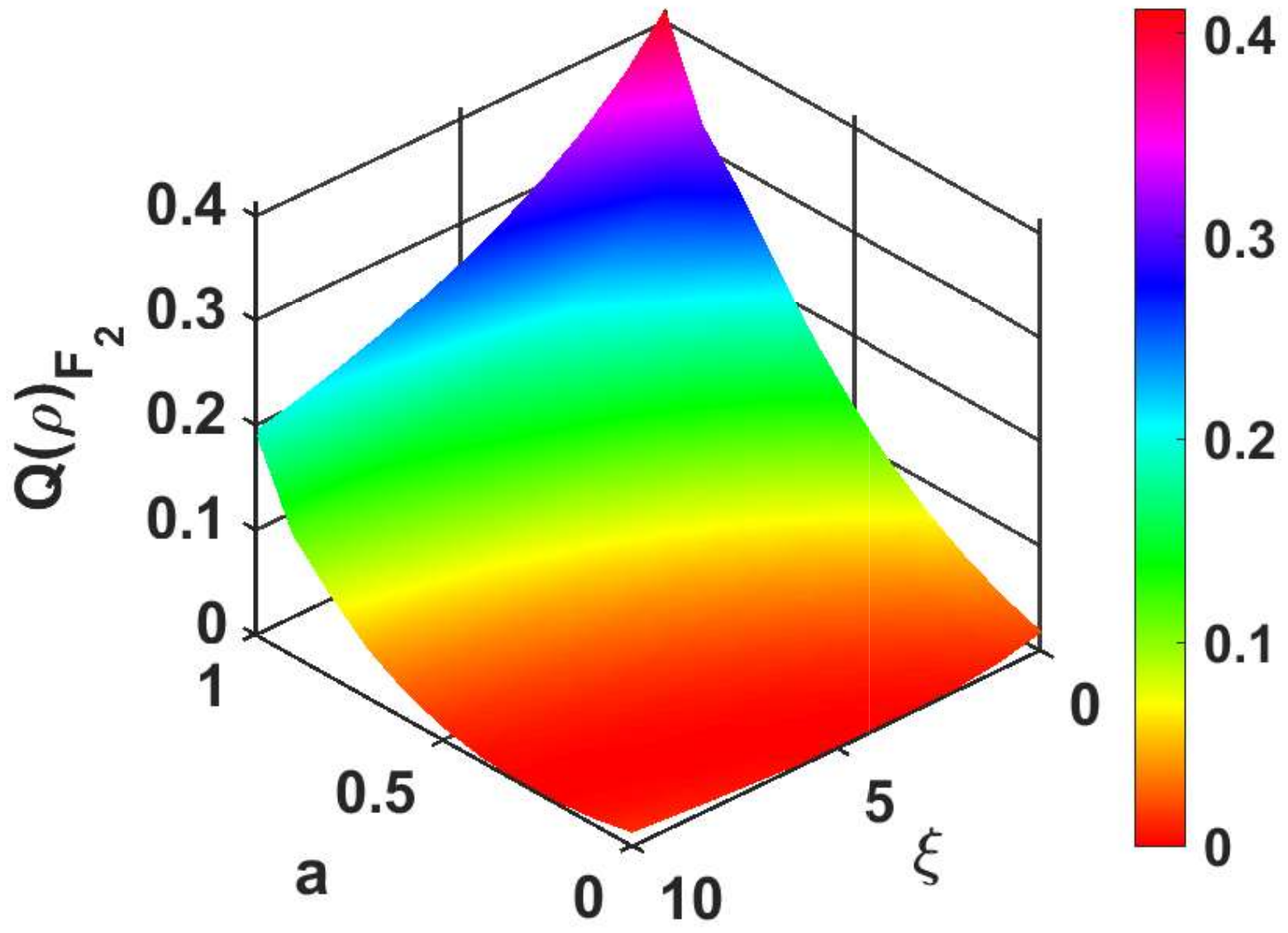}
	\caption{(Color online) Curve of stable QD with parameter $a$ and decay rate $\xi$. }
	\label{Fig6}
\end{figure}

All the above analysis is based on the fact that $\xi$ is already a definite value. However, $\xi$ can be changed by the material, production process and shape of the PWs. Therefore, we let $\xi$ a variable and plot Fig. \ref{Fig6}.

Fig. \ref{Fig6} shows the effect of $\xi$ on QD. We find that the protection effect of feedback control is best when $\xi\rightarrow 0$ and QD can be enhanced up to $0.42$. This finding offers guidance for the fabrication of PWs. By lowering the decay rate to suppress the interaction between the state and the environment, the protective ability of $F_2$ can be optimally utilized.

Again, we take Eq. (\ref{Xstate}) to be the initial state and obtain a more general stationary matrix under $F_2$:

\begin{flalign}\label{2Xrho22wentai}
	\begin{split}
	&\rho^X_{22}(t\rightarrow \infty)_{F_2}=\rho^X_{33}(t\rightarrow \infty)_{F_2}\\
	&=\frac{(12+\xi)(\mathcal{A} _3+\mathcal{B} _3)-(4+\xi)(\mathcal{C} _3+\mathcal{D} _3)+8\mathcal{E} _3}{4(8+\xi)},
	\end{split}&
\end{flalign}
\begin{flalign}\label{2Xrho23wentai}
	\begin{split}
		&\rho^X_{23}(t\rightarrow \infty)_{F_2}=\rho^X_{23}(t\rightarrow \infty)_{F_2}\\
		&=\frac{-(4+\xi)(\mathcal{A} _3+\mathcal{B} _3)+(12+\xi)(\mathcal{C} _3+\mathcal{D} _3)+8\mathcal{E} _3}{4(8+\xi)},
	\end{split}&
\end{flalign}
\begin{flalign}\label{2Xrho44wentai}
	\begin{split}
		\rho^X_{44}(t\rightarrow \infty)_{F_2}=\frac{2(4+\xi)(\mathcal{A} _3+\mathcal{B} _3+\mathcal{C} _3+\mathcal{D} _3+2\mathcal{E} _3)}{4(8+\xi)},
	\end{split}&
\end{flalign}
where $\mathcal{A} _3=\rho_{22}(0)-\rho_{11}(0)(-48+24\xi+\xi^2)/[16(12+\xi)]$, $\mathcal{B} _3=\rho_{33}(0)-\rho_{11}(0)(-48+24\xi+\xi^2)/[16(12+\xi)]$, $\mathcal{C} _3=\rho_{23}(0)-\rho_{11}(0)(-48+24\xi+\xi^2)/[16(12+\xi)]$, $\mathcal{D} _3=\rho_{32}(0)-\rho_{11}(0)(-48+24\xi+\xi^2)/[16(12+\xi)]$ and $\mathcal{E} _3=\rho_{44}(0)+\rho_{11}(0)(48+32\xi+\xi^2)/[8(12+\xi)]$.

Due to $\xi$, none of these matrix elements are equal to $0$ unless the initial matrix was originally $0$. We therefore conclude that the feedback control $F_2$ will protect against QD of all $X$ states.

\section{THE DIFFERENCE AND SIMILARITY OF FEEDBACK CONTROL}

\subsection{The Difference between $F_1$ and $F_2$}

We explore the impact of a symmetric quantum feedback control $F = \mu(\sigma_x\otimes \sigma_z + \sigma_z\otimes \sigma_x)-(\sigma_x\otimes I + I\otimes \sigma_x)$ on the quantum discord of $X$ states in the V-shaped plasmonic waveguides. $F_1=-\sigma_x\otimes \sigma_z - \sigma_z\otimes \sigma_x-\sigma_x\otimes I - I\otimes \sigma_x$ can enhance the QD of the $X$ states satisfying $ \rho_{22}(0)+\rho_{33}(0)-\rho_{23}(0)-\rho_{32}(0)\neq 0 $. In particular, for the Werner state $\rho^W(0)$, this enhancement is inversely proportional to $a$. The protective effect of $F_2=\sigma_x\otimes \sigma_z + \sigma_z\otimes \sigma_x-\sigma_x\otimes I - I\otimes \sigma_x$ is manifested for all $X$ states. Specially, for $\rho^W(0)$, this protection is proportional to $a$ and can be enhanced by decreasing the $\xi$ of plasmonic waveguides. 

In addition to the differences mentioned above, we note that the stationary matrix in Eq. (\ref{WSF1wentai}) do not contain the decay rate $\xi$, which is mathematically counterintuitive. The explanation to this problem can be found in Eqs. (\ref{f1rho22}), (\ref{f1rho23}) and (\ref{f1rho44}). As $t\rightarrow \infty$, the terms containing $\xi$ eventually vanish due to the presence of the negative exponential terms. Under $F_2$, this fact does not occur and $\xi$ is preserved. We display a more detailed analysis of this trouble in the Appendix. From the Appendix, we notice that among all the values of $\mu\in [-1,0)\cup (0,1]$, only $\mu=-1$ is an unusual point. This particular value results in a reduction of almost half of the terms in the matrix elements, which fundamentally change the evolution of the matrix element. For other values, they can only cause numerical affects. It means that we just need $F_1$ and $F_2$ to have a comprehensive understanding of $F$. That is why we only choose $\mu=-1$ and $\mu=1$.

\subsection{The Similarity and Work Mechanism of $F_1$ and $F_2$ }

Now we provide a more detailed explanation of how this feedback control works. Both feedback controls, $F_1$ and $F_2$, reduce the original $4\times 4$ matrix to a $3\times 3$ subspace. The essence of both types of feedback is to reduce the proportion of $| 1\rangle $ in quantum state. We believe that this behavior can partially restore Werner state. It has been proved that for either $| 1\rangle $ or $| 4\rangle $, their QD is always $0$ and Bell state $| \varPsi \rangle $ has maximum QD. That represents $F$ motivates the degraded Bell state, Werner state, change towards its original state. To put it another way, it divides the 4-dimensional space $\mathcal{H}$ into two parts, causing a leakage from the 1-dimensional subspace $\mathcal{H} _1$ to the 3-dimensional subspace $\mathcal{H} _2$. Mathematically, the matrix elements without feedback control consist solely of exponential decay terms, but with the addition of $F$, extra constant terms occur in the matrix elements of $\mathcal{H} _2$, which are the very reason why the stationary QD can exist. Physically, the reason QD becomes $0$ is that the quantum state interacts with environment so that energy dissipation occurs. $F$ prevents the complete dissipation of energy from $\mathcal{H} _1$ into the environment, allowing it to flow partially or completely into $\mathcal{H} _2$, or it prevents the energy in $\mathcal{H} _2$ from being completely dissipated. This explains where the protective effect of $F$ comes from. We believe that the  enhancement of QD would be more significantly enhanced if there existed some methods to reduce the weights of both $| 1\rangle $ and $| 4\rangle $.

\section{Conclusion}

We propose a feedback control that enhance the quantum discord of $X$ states. It does not change the shape of the density matrix of states. However, it regulates the fraction of different ground states in the quantum state. In the limiting case, this tuning ability will localize the matrix into $3\times 3$ subspace. As a result, energy of the system is preserved partially instead of diffusing completely into the environment. This is how the feedback mechanism works. For $X$ states, the matrix elements of the secondary diagonal are the source of quantum correlation. Therefore, the feedback Hamiltonian imposes a requirement on the initial matrix: the secondary diagonal matrix elements in the limiting case cannot be $0$. In addition, we conclude that the protective effect of the feedback control can be enhanced by decreasing the decay rate of atoms in the plasmonic waveguides.

Our work suggests the possibility of proposing an improved feedback Hamiltonian. If there exists a new feedback Hamiltonian that localizes the matrix into a smaller subspace, it may lead to a more significant enhancement of quantum discord.

\section*{ACKNOWLEDGMENTS}

This work was supported by National Natural Science Foundation of China (NSFC) under grants No.12074027.

\section*{APPENDIX: THE IMPACT OF $\mu$ ON THE DENSITY MATRIX}
\setcounter{equation}{0}
\renewcommand{\theequation}{A.\arabic{equation}}
In order to highlight the specificity of $F_1$, we ought to have calculated the equations for the   dynamical evolution of the density matrix elements without taking any particular value for $\mu$. Unfortunately, these results are too tedious to obtain effective information from them and to be presented here. However, we propose a more efficient way: qualitative analysis by means of differential equations for matrix elements. We now give the differential equation set for all matrix elements:

\begin{flalign}\label{A11}
	\begin{split}
		&\frac{\partial\rho_{11}(t)}{\partial t}=-2\rho_{11}(t)[\xi+(\mu-1)^2]\\
		&+(\mu^2-1)(\rho_{14}(t)+\rho_{41}(t))\\
		&+(\mu-1)(\rho_{22}(t)+\rho_{33}(t)+\rho_{23}(t)+\rho_{32}(t)),
	\end{split}
\end{flalign}

\begin{flalign}\label{A22}
	\begin{split}
		&\frac{\partial\rho_{22}(t)}{\partial t}=\rho_{11}(t)[\xi+(\mu-1)^2]+(\mu+1)^2\rho_{44}(t)\\
		&-(1+\mu^2)(\rho_{23}(t)+\rho_{32}(t))\\
		&-\xi(\frac{1}{2}-\frac{i}{2})(\rho_{23}(t)+i\rho_{32}(t))\\
		&-(\xi+2+2\mu^2)\rho_{22}(t)+(1-\mu^2)(\rho_{14}(t)+\rho_{41}(t))\\
		&+(\mu+1)i\sqrt{\xi}(\rho_{14}(t)+\rho_{23}(t)-\rho_{32}(t)-\rho_{41}(t)),
	\end{split}&
\end{flalign}

\begin{flalign}\label{A33}
	\begin{split}
		&\frac{\partial\rho_{33}(t)}{\partial t}=\rho_{11}(t)[\xi+(\mu-1)^2]+(\mu+1)^2\rho_{44}(t)\\
		&-(1+\mu^2)(\rho_{23}(t)+\rho_{32}(t))\\
		&-\xi(\frac{1}{2}+\frac{i}{2})(\rho_{23}(t)-i\rho_{32}(t))\\
		&-(\xi+2+2\mu^2)\rho_{33}(t)+(1-\mu^2)(\rho_{14}(t)+\rho_{41}(t))\\
		&+(\mu+1)i\sqrt{\xi}(\rho_{14}(t)-\rho_{23}(t)+\rho_{32}(t)-\rho_{41}(t)),
	\end{split}&
\end{flalign}

\begin{flalign}\label{A23}
	\begin{split}
		&\frac{\partial\rho_{23}(t)}{\partial t}=\rho_{11}(t)[\xi+(\mu-1)^2]+(\mu+1)^2\rho_{44}(t)\\
		&-(1+\mu^2)(\rho_{22}(t)+\rho_{33}(t))\\
		&-\xi(\frac{1}{2}-\frac{i}{2})(\rho_{22}(t)+i\rho_{33}(t))\\
		&-(\xi+2+2\mu^2)\rho_{23}(t)+(1-\mu^2)(\rho_{14}(t)+\rho_{41}(t))\\
		&+(\mu+1)i\sqrt{\xi}(\rho_{14}(t)+\rho_{22}(t)-\rho_{33}(t)-\rho_{41}(t)),
	\end{split}&
\end{flalign}

\begin{flalign}\label{A32}
	\begin{split}
		&\frac{\partial\rho_{32}(t)}{\partial t}=\rho_{11}(t)[\xi+(\mu-1)^2]+(\mu+1)^2\rho_{44}(t)\\
		&-(1+\mu^2)(\rho_{22}(t)+\rho_{33}(t))\\
		&-\xi(\frac{1}{2}+\frac{i}{2})(\rho_{22}(t)-i\rho_{33}(t))\\
		&-(\xi+2+2\mu^2)\rho_{32}(t)+(1-\mu^2)(\rho_{14}(t)+\rho_{41}(t))\\
		&+(\mu+1)i\sqrt{\xi}(\rho_{14}(t)-\rho_{22}(t)+\rho_{33}(t)-\rho_{41}(t)),
	\end{split}&
\end{flalign}

\begin{flalign}\label{A44}
	\begin{split}
		&\frac{\partial\rho_{44}(t)}{\partial t}=(\mu+1)2i\sqrt{\xi}(\rho_{14}(t)+\rho_{41}(t))\\
		&+(\mu^2-1)(\rho_{14}(t)+\rho_{41}(t))-2(\mu+1)^2\rho_{44}(t)\\
		&+[\xi+(\mu+1)^2](\rho_{22}(t)+\rho_{33}(t)+\rho_{23}(t)+\rho_{32}(t)).
	\end{split}&
\end{flalign}

In the above equations, there are several terms contain $(\mu+1)$. When $\mu = -1$, these terms vanish which leads to Eqs.(\ref{f1rho11}), (\ref{f1rho22}), (\ref{f1rho23}) and (\ref{f1rho44}). However, this situation does not happen when $\mu$ takes other values. With this simple comparison, we believe that $F_1$ and $F_2$ are enough to have a complete realization of feedback control $F$.

\bibliography{Primary_manuscript}

\begin{thebibliography}{52}%
\makeatletter
\providecommand \@ifxundefined [1]{%
 \@ifx{#1\undefined}
}%
\providecommand \@ifnum [1]{%
 \ifnum #1\expandafter \@firstoftwo
 \else \expandafter \@secondoftwo
 \fi
}%
\providecommand \@ifx [1]{%
 \ifx #1\expandafter \@firstoftwo
 \else \expandafter \@secondoftwo
 \fi
}%
\providecommand \natexlab [1]{#1}%
\providecommand \enquote  [1]{``#1''}%
\providecommand \bibnamefont  [1]{#1}%
\providecommand \bibfnamefont [1]{#1}%
\providecommand \citenamefont [1]{#1}%
\providecommand \href@noop [0]{\@secondoftwo}%
\providecommand \href [0]{\begingroup \@sanitize@url \@href}%
\providecommand \@href[1]{\@@startlink{#1}\@@href}%
\providecommand \@@href[1]{\endgroup#1\@@endlink}%
\providecommand \@sanitize@url [0]{\catcode `\\12\catcode `\$12\catcode `\&12\catcode `\#12\catcode `\^12\catcode `\_12\catcode `\%12\relax}%
\providecommand \@@startlink[1]{}%
\providecommand \@@endlink[0]{}%
\providecommand \url  [0]{\begingroup\@sanitize@url \@url }%
\providecommand \@url [1]{\endgroup\@href {#1}{\urlprefix }}%
\providecommand \urlprefix  [0]{URL }%
\providecommand \Eprint [0]{\href }%
\providecommand \doibase [0]{http://dx.doi.org/}%
\providecommand \selectlanguage [0]{\@gobble}%
\providecommand \bibinfo  [0]{\@secondoftwo}%
\providecommand \bibfield  [0]{\@secondoftwo}%
\providecommand \translation [1]{[#1]}%
\providecommand \BibitemOpen [0]{}%
\providecommand \bibitemStop [0]{}%
\providecommand \bibitemNoStop [0]{.\EOS\space}%
\providecommand \EOS [0]{\spacefactor3000\relax}%
\providecommand \BibitemShut  [1]{\csname bibitem#1\endcsname}%
\let\auto@bib@innerbib\@empty
%</preamble>
\bibitem [{\citenamefont {Luo}\ \emph {et~al.}(2019)\citenamefont {Luo}, \citenamefont {Zhong}, \citenamefont {Erhard}, \citenamefont {Wang}, \citenamefont {Peng}, \citenamefont {Krenn}, \citenamefont {Jiang}, \citenamefont {Li}, \citenamefont {Liu}, \citenamefont {Lu}, \citenamefont {Zeilinger},\ and\ \citenamefont {Pan}}]{PhysRevLett.123.070505}%
  \BibitemOpen
  \bibfield  {author} {\bibinfo {author} {\bibfnamefont {Y.~H.}\ \bibnamefont {Luo}}, \bibinfo {author} {\bibfnamefont {H.~S.}\ \bibnamefont {Zhong}}, \bibinfo {author} {\bibfnamefont {M.}~\bibnamefont {Erhard}}, \bibinfo {author} {\bibfnamefont {X.~L.}\ \bibnamefont {Wang}}, \bibinfo {author} {\bibfnamefont {L.~C.}\ \bibnamefont {Peng}}, \bibinfo {author} {\bibfnamefont {M.}~\bibnamefont {Krenn}}, \bibinfo {author} {\bibfnamefont {X.}~\bibnamefont {Jiang}}, \bibinfo {author} {\bibfnamefont {L.}~\bibnamefont {Li}}, \bibinfo {author} {\bibfnamefont {N.~L.}\ \bibnamefont {Liu}}, \bibinfo {author} {\bibfnamefont {C.~Y.}\ \bibnamefont {Lu}}, \bibinfo {author} {\bibfnamefont {A.}~\bibnamefont {Zeilinger}}, \ and\ \bibinfo {author} {\bibfnamefont {J.~W.}\ \bibnamefont {Pan}},\ }\bibfield  {title} {\enquote {\bibinfo {title} {Quantum teleportation in high dimensions},}\ }\href {\doibase 10.1103/PhysRevLett.123.070505} {\bibfield  {journal} {\bibinfo  {journal} {Phys. Rev. Lett.}\ }\textbf {\bibinfo {volume} {123}},\ \bibinfo {pages} {070505} (\bibinfo {year} {2019})}\BibitemShut {NoStop}%
\bibitem [{\citenamefont {Wu}\ \emph {et~al.}(2022)\citenamefont {Wu}, \citenamefont {Long},\ and\ \citenamefont {Hayashi}}]{PhysRevApplied.17.064011}%
  \BibitemOpen
  \bibfield  {author} {\bibinfo {author} {\bibfnamefont {J.~W.}\ \bibnamefont {Wu}}, \bibinfo {author} {\bibfnamefont {G.~L.}\ \bibnamefont {Long}}, \ and\ \bibinfo {author} {\bibfnamefont {M.}~\bibnamefont {Hayashi}},\ }\bibfield  {title} {\enquote {\bibinfo {title} {Quantum secure direct communication with private dense coding using a general preshared quantum state},}\ }\href {\doibase 10.1103/PhysRevApplied.17.064011} {\bibfield  {journal} {\bibinfo  {journal} {Phys. Rev. Appl.}\ }\textbf {\bibinfo {volume} {17}},\ \bibinfo {pages} {064011} (\bibinfo {year} {2022})}\BibitemShut {NoStop}%
\bibitem [{\citenamefont {Tordrup}\ \emph {et~al.}(2008)\citenamefont {Tordrup}, \citenamefont {Negretti},\ and\ \citenamefont {M\o{}lmer}}]{PhysRevLett.101.040501}%
  \BibitemOpen
  \bibfield  {author} {\bibinfo {author} {\bibfnamefont {K.}~\bibnamefont {Tordrup}}, \bibinfo {author} {\bibfnamefont {A.}~\bibnamefont {Negretti}}, \ and\ \bibinfo {author} {\bibfnamefont {K.}~\bibnamefont {M\o{}lmer}},\ }\bibfield  {title} {\enquote {\bibinfo {title} {Holographic quantum computing},}\ }\href {\doibase 10.1103/PhysRevLett.101.040501} {\bibfield  {journal} {\bibinfo  {journal} {Phys. Rev. Lett.}\ }\textbf {\bibinfo {volume} {101}},\ \bibinfo {pages} {040501} (\bibinfo {year} {2008})}\BibitemShut {NoStop}%
\bibitem [{\citenamefont {Hutter}\ and\ \citenamefont {Loss}(2016)}]{PhysRevB.93.125105}%
  \BibitemOpen
  \bibfield  {author} {\bibinfo {author} {\bibfnamefont {A.}~\bibnamefont {Hutter}}\ and\ \bibinfo {author} {\bibfnamefont {D.}~\bibnamefont {Loss}},\ }\bibfield  {title} {\enquote {\bibinfo {title} {Quantum computing with parafermions},}\ }\href {\doibase 10.1103/PhysRevB.93.125105} {\bibfield  {journal} {\bibinfo  {journal} {Phys. Rev. B}\ }\textbf {\bibinfo {volume} {93}},\ \bibinfo {pages} {125105} (\bibinfo {year} {2016})}\BibitemShut {NoStop}%
\bibitem [{\citenamefont {Vandersypen}\ \emph {et~al.}(2001)\citenamefont {Vandersypen}, \citenamefont {Steffen}, \citenamefont {Breyta}, \citenamefont {Yannoni}, \citenamefont {Sherwood},\ and\ \citenamefont {Chuang}}]{Vandersypen_2001}%
  \BibitemOpen
  \bibfield  {author} {\bibinfo {author} {\bibfnamefont {L.~M.~K.}\ \bibnamefont {Vandersypen}}, \bibinfo {author} {\bibfnamefont {M.}~\bibnamefont {Steffen}}, \bibinfo {author} {\bibfnamefont {G.}~\bibnamefont {Breyta}}, \bibinfo {author} {\bibfnamefont {C.~S.}\ \bibnamefont {Yannoni}}, \bibinfo {author} {\bibfnamefont {M.~H.}\ \bibnamefont {Sherwood}}, \ and\ \bibinfo {author} {\bibfnamefont {I.~L.}\ \bibnamefont {Chuang}},\ }\bibfield  {title} {\enquote {\bibinfo {title} {Experimental realization of shor{\textquotesingle}s quantum factoring algorithm using nuclear magnetic resonance},}\ }\href {\doibase 10.1038/414883a} {\bibfield  {journal} {\bibinfo  {journal} {Nature}\ }\textbf {\bibinfo {volume} {414}},\ \bibinfo {pages} {883--887} (\bibinfo {year} {2001})}\BibitemShut {NoStop}%
\bibitem [{\citenamefont {Braunstein}\ \emph {et~al.}(1999)\citenamefont {Braunstein}, \citenamefont {Caves}, \citenamefont {Jozsa}, \citenamefont {Linden}, \citenamefont {Popescu},\ and\ \citenamefont {Schack}}]{PhysRevLett.83.1054}%
  \BibitemOpen
  \bibfield  {author} {\bibinfo {author} {\bibfnamefont {S.~L.}\ \bibnamefont {Braunstein}}, \bibinfo {author} {\bibfnamefont {C.~M.}\ \bibnamefont {Caves}}, \bibinfo {author} {\bibfnamefont {R.}~\bibnamefont {Jozsa}}, \bibinfo {author} {\bibfnamefont {N.}~\bibnamefont {Linden}}, \bibinfo {author} {\bibfnamefont {S.}~\bibnamefont {Popescu}}, \ and\ \bibinfo {author} {\bibfnamefont {R.}~\bibnamefont {Schack}},\ }\bibfield  {title} {\enquote {\bibinfo {title} {Separability of very noisy mixed states and implications for nmr quantum computing},}\ }\href {\doibase 10.1103/PhysRevLett.83.1054} {\bibfield  {journal} {\bibinfo  {journal} {Phys. Rev. Lett.}\ }\textbf {\bibinfo {volume} {83}},\ \bibinfo {pages} {1054--1057} (\bibinfo {year} {1999})}\BibitemShut {NoStop}%
\bibitem [{\citenamefont {Du}\ \emph {et~al.}(2001)\citenamefont {Du}, \citenamefont {Shi}, \citenamefont {Zhou}, \citenamefont {Fan}, \citenamefont {Ye}, \citenamefont {Han},\ and\ \citenamefont {Wu}}]{PhysRevA.64.042306}%
  \BibitemOpen
  \bibfield  {author} {\bibinfo {author} {\bibfnamefont {J.~F.}\ \bibnamefont {Du}}, \bibinfo {author} {\bibfnamefont {M.~J.}\ \bibnamefont {Shi}}, \bibinfo {author} {\bibfnamefont {X.~Y.}\ \bibnamefont {Zhou}}, \bibinfo {author} {\bibfnamefont {Y.~M.}\ \bibnamefont {Fan}}, \bibinfo {author} {\bibfnamefont {B.~J.}\ \bibnamefont {Ye}}, \bibinfo {author} {\bibfnamefont {R.~D.}\ \bibnamefont {Han}}, \ and\ \bibinfo {author} {\bibfnamefont {J.~H.}\ \bibnamefont {Wu}},\ }\bibfield  {title} {\enquote {\bibinfo {title} {Implementation of a quantum algorithm to solve the bernstein-vazirani parity problem without entanglement on an ensemble quantum computer},}\ }\href {\doibase 10.1103/PhysRevA.64.042306} {\bibfield  {journal} {\bibinfo  {journal} {Phys. Rev. A}\ }\textbf {\bibinfo {volume} {64}},\ \bibinfo {pages} {042306} (\bibinfo {year} {2001})}\BibitemShut {NoStop}%
\bibitem [{\citenamefont {Datta}\ \emph {et~al.}(2005)\citenamefont {Datta}, \citenamefont {Flammia},\ and\ \citenamefont {Caves}}]{PhysRevA.72.042316}%
  \BibitemOpen
  \bibfield  {author} {\bibinfo {author} {\bibfnamefont {A.}~\bibnamefont {Datta}}, \bibinfo {author} {\bibfnamefont {S.~T.}\ \bibnamefont {Flammia}}, \ and\ \bibinfo {author} {\bibfnamefont {C.~M.}\ \bibnamefont {Caves}},\ }\bibfield  {title} {\enquote {\bibinfo {title} {Entanglement and the power of one qubit},}\ }\href {\doibase 10.1103/PhysRevA.72.042316} {\bibfield  {journal} {\bibinfo  {journal} {Phys. Rev. A}\ }\textbf {\bibinfo {volume} {72}},\ \bibinfo {pages} {042316} (\bibinfo {year} {2005})}\BibitemShut {NoStop}%
\bibitem [{\citenamefont {Meyer}(2000)}]{PhysRevLett.85.2014}%
  \BibitemOpen
  \bibfield  {author} {\bibinfo {author} {\bibfnamefont {D.~A.}\ \bibnamefont {Meyer}},\ }\bibfield  {title} {\enquote {\bibinfo {title} {Sophisticated quantum search without entanglement},}\ }\href {\doibase 10.1103/PhysRevLett.85.2014} {\bibfield  {journal} {\bibinfo  {journal} {Phys. Rev. Lett.}\ }\textbf {\bibinfo {volume} {85}},\ \bibinfo {pages} {2014--2017} (\bibinfo {year} {2000})}\BibitemShut {NoStop}%
\bibitem [{\citenamefont {Lanyon}\ \emph {et~al.}(2008)\citenamefont {Lanyon}, \citenamefont {Barbieri}, \citenamefont {Almeida},\ and\ \citenamefont {White}}]{PhysRevLett.101.200501}%
  \BibitemOpen
  \bibfield  {author} {\bibinfo {author} {\bibfnamefont {B.~P.}\ \bibnamefont {Lanyon}}, \bibinfo {author} {\bibfnamefont {M.}~\bibnamefont {Barbieri}}, \bibinfo {author} {\bibfnamefont {M.~P.}\ \bibnamefont {Almeida}}, \ and\ \bibinfo {author} {\bibfnamefont {A.~G.}\ \bibnamefont {White}},\ }\bibfield  {title} {\enquote {\bibinfo {title} {Experimental quantum computing without entanglement},}\ }\href {\doibase 10.1103/PhysRevLett.101.200501} {\bibfield  {journal} {\bibinfo  {journal} {Phys. Rev. Lett.}\ }\textbf {\bibinfo {volume} {101}},\ \bibinfo {pages} {200501} (\bibinfo {year} {2008})}\BibitemShut {NoStop}%
\bibitem [{\citenamefont {Groisman}\ \emph {et~al.}(2005)\citenamefont {Groisman}, \citenamefont {Popescu},\ and\ \citenamefont {Winter}}]{PhysRevA.72.032317}%
  \BibitemOpen
  \bibfield  {author} {\bibinfo {author} {\bibfnamefont {B.}~\bibnamefont {Groisman}}, \bibinfo {author} {\bibfnamefont {S.}~\bibnamefont {Popescu}}, \ and\ \bibinfo {author} {\bibfnamefont {A.}~\bibnamefont {Winter}},\ }\bibfield  {title} {\enquote {\bibinfo {title} {Quantum, classical, and total amount of correlations in a quantum state},}\ }\href {\doibase 10.1103/PhysRevA.72.032317} {\bibfield  {journal} {\bibinfo  {journal} {Phys. Rev. A}\ }\textbf {\bibinfo {volume} {72}},\ \bibinfo {pages} {032317} (\bibinfo {year} {2005})}\BibitemShut {NoStop}%
\bibitem [{\citenamefont {Ollivier}\ and\ \citenamefont {Zurek}(2001)}]{PhysRevLett.88.017901}%
  \BibitemOpen
  \bibfield  {author} {\bibinfo {author} {\bibfnamefont {H.}~\bibnamefont {Ollivier}}\ and\ \bibinfo {author} {\bibfnamefont {W.~H.}\ \bibnamefont {Zurek}},\ }\bibfield  {title} {\enquote {\bibinfo {title} {Quantum discord: A measure of the quantumness of correlations},}\ }\href {\doibase 10.1103/PhysRevLett.88.017901} {\bibfield  {journal} {\bibinfo  {journal} {Phys. Rev. Lett.}\ }\textbf {\bibinfo {volume} {88}},\ \bibinfo {pages} {017901} (\bibinfo {year} {2001})}\BibitemShut {NoStop}%
\bibitem [{\citenamefont {Henderson}\ and\ \citenamefont {Vedral}(2001)}]{LHenderson_2001}%
  \BibitemOpen
  \bibfield  {author} {\bibinfo {author} {\bibfnamefont {L.}~\bibnamefont {Henderson}}\ and\ \bibinfo {author} {\bibfnamefont {V.}~\bibnamefont {Vedral}},\ }\bibfield  {title} {\enquote {\bibinfo {title} {Classical, quantum and total correlations},}\ }\href {\doibase 10.1088/0305-4470/34/35/315} {\bibfield  {journal} {\bibinfo  {journal} {Journal of Physics A: Mathematical and General}\ }\textbf {\bibinfo {volume} {34}},\ \bibinfo {pages} {6899} (\bibinfo {year} {2001})}\BibitemShut {NoStop}%
\bibitem [{\citenamefont {Maziero}\ \emph {et~al.}(2009)\citenamefont {Maziero}, \citenamefont {C\'eleri}, \citenamefont {Serra},\ and\ \citenamefont {Vedral}}]{PhysRevA.80.044102}%
  \BibitemOpen
  \bibfield  {author} {\bibinfo {author} {\bibfnamefont {J.}~\bibnamefont {Maziero}}, \bibinfo {author} {\bibfnamefont {L.~C.}\ \bibnamefont {C\'eleri}}, \bibinfo {author} {\bibfnamefont {R.~M.}\ \bibnamefont {Serra}}, \ and\ \bibinfo {author} {\bibfnamefont {V.}~\bibnamefont {Vedral}},\ }\bibfield  {title} {\enquote {\bibinfo {title} {Classical and quantum correlations under decoherence},}\ }\href {\doibase 10.1103/PhysRevA.80.044102} {\bibfield  {journal} {\bibinfo  {journal} {Phys. Rev. A}\ }\textbf {\bibinfo {volume} {80}},\ \bibinfo {pages} {044102} (\bibinfo {year} {2009})}\BibitemShut {NoStop}%
\bibitem [{\citenamefont {Luo}(2008)}]{PhysRevA.77.042303}%
  \BibitemOpen
  \bibfield  {author} {\bibinfo {author} {\bibfnamefont {S.~L.}\ \bibnamefont {Luo}},\ }\bibfield  {title} {\enquote {\bibinfo {title} {Quantum discord for two-qubit systems},}\ }\href {\doibase 10.1103/PhysRevA.77.042303} {\bibfield  {journal} {\bibinfo  {journal} {Phys. Rev. A}\ }\textbf {\bibinfo {volume} {77}},\ \bibinfo {pages} {042303} (\bibinfo {year} {2008})}\BibitemShut {NoStop}%
\bibitem [{\citenamefont {Carvalho}\ and\ \citenamefont {Hope}(2007)}]{PhysRevA.76.010301}%
  \BibitemOpen
  \bibfield  {author} {\bibinfo {author} {\bibfnamefont {A.~R.~R.}\ \bibnamefont {Carvalho}}\ and\ \bibinfo {author} {\bibfnamefont {J.~J.}\ \bibnamefont {Hope}},\ }\bibfield  {title} {\enquote {\bibinfo {title} {Stabilizing entanglement by quantum-jump-based feedback},}\ }\href {\doibase 10.1103/PhysRevA.76.010301} {\bibfield  {journal} {\bibinfo  {journal} {Phys. Rev. A}\ }\textbf {\bibinfo {volume} {76}},\ \bibinfo {pages} {010301(R)} (\bibinfo {year} {2007})}\BibitemShut {NoStop}%
\bibitem [{\citenamefont {Zou}\ \emph {et~al.}(2017)\citenamefont {Zou}, \citenamefont {Li}, \citenamefont {Wang}, \citenamefont {Cao}, \citenamefont {Ren}, \citenamefont {Yin}, \citenamefont {Peng}, \citenamefont {Wang},\ and\ \citenamefont {Pan}}]{PhysRevA.95.042342}%
  \BibitemOpen
  \bibfield  {author} {\bibinfo {author} {\bibfnamefont {W.~J.}\ \bibnamefont {Zou}}, \bibinfo {author} {\bibfnamefont {Y.~H.}\ \bibnamefont {Li}}, \bibinfo {author} {\bibfnamefont {S.~C.}\ \bibnamefont {Wang}}, \bibinfo {author} {\bibfnamefont {Y.}~\bibnamefont {Cao}}, \bibinfo {author} {\bibfnamefont {J.~G.}\ \bibnamefont {Ren}}, \bibinfo {author} {\bibfnamefont {J.}~\bibnamefont {Yin}}, \bibinfo {author} {\bibfnamefont {C.~Z.}\ \bibnamefont {Peng}}, \bibinfo {author} {\bibfnamefont {X.~B.}\ \bibnamefont {Wang}}, \ and\ \bibinfo {author} {\bibfnamefont {J.~W.}\ \bibnamefont {Pan}},\ }\bibfield  {title} {\enquote {\bibinfo {title} {Protecting entanglement from finite-temperature thermal noise via weak measurement and quantum measurement reversal},}\ }\href {\doibase 10.1103/PhysRevA.95.042342} {\bibfield  {journal} {\bibinfo  {journal} {Phys. Rev. A}\ }\textbf {\bibinfo {volume} {95}},\ \bibinfo {pages} {042342} (\bibinfo {year} {2017})}\BibitemShut {NoStop}%
\bibitem [{\citenamefont {Bozhevolnyi}\ \emph {et~al.}(2006)\citenamefont {Bozhevolnyi}, \citenamefont {Volkov}, \citenamefont {Devaux}, \citenamefont {Laluet},\ and\ \citenamefont {Ebbesen}}]{Bozhevolnyi2006ChannelPS}%
  \BibitemOpen
  \bibfield  {author} {\bibinfo {author} {\bibfnamefont {S.~I.}\ \bibnamefont {Bozhevolnyi}}, \bibinfo {author} {\bibfnamefont {V.~S.}\ \bibnamefont {Volkov}}, \bibinfo {author} {\bibfnamefont {{\'E}.}~\bibnamefont {Devaux}}, \bibinfo {author} {\bibfnamefont {J.-Y.}\ \bibnamefont {Laluet}}, \ and\ \bibinfo {author} {\bibfnamefont {T.~W.}\ \bibnamefont {Ebbesen}},\ }\bibfield  {title} {\enquote {\bibinfo {title} {Channel plasmon subwavelength waveguide components including interferometers and ring resonators},}\ }\href {https://api.semanticscholar.org/CorpusID:4426568} {\bibfield  {journal} {\bibinfo  {journal} {Nature}\ }\textbf {\bibinfo {volume} {440}},\ \bibinfo {pages} {508--511} (\bibinfo {year} {2006})}\BibitemShut {NoStop}%
\bibitem [{\citenamefont {Brongersma}\ and\ \citenamefont {Shalaev}(2010)}]{Brongersma2010TheCF}%
  \BibitemOpen
  \bibfield  {author} {\bibinfo {author} {\bibfnamefont {M.~L.}\ \bibnamefont {Brongersma}}\ and\ \bibinfo {author} {\bibfnamefont {V.~M.}\ \bibnamefont {Shalaev}},\ }\bibfield  {title} {\enquote {\bibinfo {title} {The case for plasmonics},}\ }\href {https://api.semanticscholar.org/CorpusID:206525334} {\bibfield  {journal} {\bibinfo  {journal} {Science}\ }\textbf {\bibinfo {volume} {328}},\ \bibinfo {pages} {440 -- 441} (\bibinfo {year} {2010})}\BibitemShut {NoStop}%
\bibitem [{\citenamefont {Schuller}\ \emph {et~al.}(2010)\citenamefont {Schuller}, \citenamefont {Barnard}, \citenamefont {Cai}, \citenamefont {Jun}, \citenamefont {White},\ and\ \citenamefont {Brongersma}}]{article}%
  \BibitemOpen
  \bibfield  {author} {\bibinfo {author} {\bibfnamefont {J.}~\bibnamefont {Schuller}}, \bibinfo {author} {\bibfnamefont {E.}~\bibnamefont {Barnard}}, \bibinfo {author} {\bibfnamefont {W.~S.}\ \bibnamefont {Cai}}, \bibinfo {author} {\bibfnamefont {Y.}~\bibnamefont {Jun}}, \bibinfo {author} {\bibfnamefont {J.}~\bibnamefont {White}}, \ and\ \bibinfo {author} {\bibfnamefont {M.}~\bibnamefont {Brongersma}},\ }\bibfield  {title} {\enquote {\bibinfo {title} {Plasmonics for extreme light concentration and manipulation},}\ }\href {\doibase 10.1038/nmat2630} {\bibfield  {journal} {\bibinfo  {journal} {Nature Materials}\ }\textbf {\bibinfo {volume} {9}},\ \bibinfo {pages} {193--204} (\bibinfo {year} {2010})}\BibitemShut {NoStop}%
\bibitem [{\citenamefont {Mart\'{\i}n-Cano}\ \emph {et~al.}(2011)\citenamefont {Mart\'{\i}n-Cano}, \citenamefont {Gonz\'alez-Tudela}, \citenamefont {Mart\'{\i}n-Moreno}, \citenamefont {Garc\'{\i}a-Vidal}, \citenamefont {Tejedor},\ and\ \citenamefont {Moreno}}]{PhysRevB.84.235306}%
  \BibitemOpen
  \bibfield  {author} {\bibinfo {author} {\bibfnamefont {D.}~\bibnamefont {Mart\'{\i}n-Cano}}, \bibinfo {author} {\bibfnamefont {A.}~\bibnamefont {Gonz\'alez-Tudela}}, \bibinfo {author} {\bibfnamefont {L.}~\bibnamefont {Mart\'{\i}n-Moreno}}, \bibinfo {author} {\bibfnamefont {F.~J.}\ \bibnamefont {Garc\'{\i}a-Vidal}}, \bibinfo {author} {\bibfnamefont {C.}~\bibnamefont {Tejedor}}, \ and\ \bibinfo {author} {\bibfnamefont {E.}~\bibnamefont {Moreno}},\ }\bibfield  {title} {\enquote {\bibinfo {title} {Dissipation-driven generation of two-qubit entanglement mediated by plasmonic waveguides},}\ }\href {\doibase 10.1103/PhysRevB.84.235306} {\bibfield  {journal} {\bibinfo  {journal} {Phys. Rev. B}\ }\textbf {\bibinfo {volume} {84}},\ \bibinfo {pages} {235306} (\bibinfo {year} {2011})}\BibitemShut {NoStop}%
\bibitem [{\citenamefont {Bouhelier}\ and\ \citenamefont {Wiederrecht}(2005)}]{PhysRevB.71.195406}%
  \BibitemOpen
  \bibfield  {author} {\bibinfo {author} {\bibfnamefont {A.}~\bibnamefont {Bouhelier}}\ and\ \bibinfo {author} {\bibfnamefont {G.~P.}\ \bibnamefont {Wiederrecht}},\ }\bibfield  {title} {\enquote {\bibinfo {title} {Excitation of broadband surface plasmon polaritons: Plasmonic continuum spectroscopy},}\ }\href {\doibase 10.1103/PhysRevB.71.195406} {\bibfield  {journal} {\bibinfo  {journal} {Phys. Rev. B}\ }\textbf {\bibinfo {volume} {71}},\ \bibinfo {pages} {195406} (\bibinfo {year} {2005})}\BibitemShut {NoStop}%
\bibitem [{\citenamefont {Hou}\ \emph {et~al.}(2010)\citenamefont {Hou}, \citenamefont {Huang},\ and\ \citenamefont {Yi}}]{PhysRevA.82.012336}%
  \BibitemOpen
  \bibfield  {author} {\bibinfo {author} {\bibfnamefont {S.~C.}\ \bibnamefont {Hou}}, \bibinfo {author} {\bibfnamefont {X.~L.}\ \bibnamefont {Huang}}, \ and\ \bibinfo {author} {\bibfnamefont {X.~X.}\ \bibnamefont {Yi}},\ }\bibfield  {title} {\enquote {\bibinfo {title} {Suppressing decoherence and improving entanglement by quantum-jump-based feedback control in two-level systems},}\ }\href {\doibase 10.1103/PhysRevA.82.012336} {\bibfield  {journal} {\bibinfo  {journal} {Phys. Rev. A}\ }\textbf {\bibinfo {volume} {82}},\ \bibinfo {pages} {012336} (\bibinfo {year} {2010})}\BibitemShut {NoStop}%
\bibitem [{\citenamefont {Wiseman}(1994)}]{PhysRevA.49.2133}%
  \BibitemOpen
  \bibfield  {author} {\bibinfo {author} {\bibfnamefont {H.~M.}\ \bibnamefont {Wiseman}},\ }\bibfield  {title} {\enquote {\bibinfo {title} {Quantum theory of continuous feedback},}\ }\href {\doibase 10.1103/PhysRevA.49.2133} {\bibfield  {journal} {\bibinfo  {journal} {Phys. Rev. A}\ }\textbf {\bibinfo {volume} {49}},\ \bibinfo {pages} {2133--2150} (\bibinfo {year} {1994})}\BibitemShut {NoStop}%
\bibitem [{\citenamefont {Jiang}\ \emph {et~al.}(2018)\citenamefont {Jiang}, \citenamefont {Wu},\ and\ \citenamefont {Yang}}]{PhysRevA.98.052134}%
  \BibitemOpen
  \bibfield  {author} {\bibinfo {author} {\bibfnamefont {W.}~\bibnamefont {Jiang}}, \bibinfo {author} {\bibfnamefont {F.~Z.}\ \bibnamefont {Wu}}, \ and\ \bibinfo {author} {\bibfnamefont {G.~J.}\ \bibnamefont {Yang}},\ }\bibfield  {title} {\enquote {\bibinfo {title} {Non-markovian entanglement dynamics of open quantum systems with continuous measurement feedback},}\ }\href {\doibase 10.1103/PhysRevA.98.052134} {\bibfield  {journal} {\bibinfo  {journal} {Phys. Rev. A}\ }\textbf {\bibinfo {volume} {98}},\ \bibinfo {pages} {052134} (\bibinfo {year} {2018})}\BibitemShut {NoStop}%
\bibitem [{\citenamefont {Xie}\ and\ \citenamefont {Xu}(2019)}]{Xie2019EnhancingPO}%
  \BibitemOpen
  \bibfield  {author} {\bibinfo {author} {\bibfnamefont {D.}~\bibnamefont {Xie}}\ and\ \bibinfo {author} {\bibfnamefont {C.~L.}\ \bibnamefont {Xu}},\ }\bibfield  {title} {\enquote {\bibinfo {title} {Enhancing precision of damping rate by pt symmetric hamiltonian},}\ }\href {https://api.semanticscholar.org/CorpusID:254987778} {\bibfield  {journal} {\bibinfo  {journal} {Quantum Information Processing}\ }\textbf {\bibinfo {volume} {18}} (\bibinfo {year} {2019})}\BibitemShut {NoStop}%
\bibitem [{\citenamefont {de~Vega}\ and\ \citenamefont {Alonso}(2017)}]{RevModPhys.89.015001}%
  \BibitemOpen
  \bibfield  {author} {\bibinfo {author} {\bibfnamefont {I.}~\bibnamefont {de~Vega}}\ and\ \bibinfo {author} {\bibfnamefont {D.}~\bibnamefont {Alonso}},\ }\bibfield  {title} {\enquote {\bibinfo {title} {Dynamics of non-markovian open quantum systems},}\ }\href {\doibase 10.1103/RevModPhys.89.015001} {\bibfield  {journal} {\bibinfo  {journal} {Rev. Mod. Phys.}\ }\textbf {\bibinfo {volume} {89}},\ \bibinfo {pages} {015001} (\bibinfo {year} {2017})}\BibitemShut {NoStop}%
\bibitem [{\citenamefont {Yamamoto}(2005)}]{PhysRevA.72.024104}%
  \BibitemOpen
  \bibfield  {author} {\bibinfo {author} {\bibfnamefont {N.}~\bibnamefont {Yamamoto}},\ }\bibfield  {title} {\enquote {\bibinfo {title} {Parametrization of the feedback hamiltonian realizing a pure steady state},}\ }\href {\doibase 10.1103/PhysRevA.72.024104} {\bibfield  {journal} {\bibinfo  {journal} {Phys. Rev. A}\ }\textbf {\bibinfo {volume} {72}},\ \bibinfo {pages} {024104} (\bibinfo {year} {2005})}\BibitemShut {NoStop}%
\bibitem [{\citenamefont {Hohenester}\ and\ \citenamefont {Truegler}(2008)}]{Hohenester2008InteractionOS}%
  \BibitemOpen
  \bibfield  {author} {\bibinfo {author} {\bibfnamefont {U.}~\bibnamefont {Hohenester}}\ and\ \bibinfo {author} {\bibfnamefont {A.}~\bibnamefont {Truegler}},\ }\bibfield  {title} {\enquote {\bibinfo {title} {Interaction of single molecules with metallic nanoparticles},}\ }\href {https://api.semanticscholar.org/CorpusID:44845545} {\bibfield  {journal} {\bibinfo  {journal} {IEEE Journal of Selected Topics in Quantum Electronics}\ }\textbf {\bibinfo {volume} {14}},\ \bibinfo {pages} {1430--1440} (\bibinfo {year} {2008})}\BibitemShut {NoStop}%
\bibitem [{\citenamefont {Mart{\'i}n-Cano}\ \emph {et~al.}(2010)\citenamefont {Mart{\'i}n-Cano}, \citenamefont {Mart{\'i}n-Moreno}, \citenamefont {Garc{\'i}a-Vidal},\ and\ \citenamefont {Moreno}}]{MartnCano2010ResonanceET}%
  \BibitemOpen
  \bibfield  {author} {\bibinfo {author} {\bibfnamefont {D.}~\bibnamefont {Mart{\'i}n-Cano}}, \bibinfo {author} {\bibfnamefont {L.}~\bibnamefont {Mart{\'i}n-Moreno}}, \bibinfo {author} {\bibfnamefont {F.~J.}\ \bibnamefont {Garc{\'i}a-Vidal}}, \ and\ \bibinfo {author} {\bibfnamefont {E.}~\bibnamefont {Moreno}},\ }\bibfield  {title} {\enquote {\bibinfo {title} {Resonance energy transfer and superradiance mediated by plasmonic nanowaveguides.}}\ }\href {https://api.semanticscholar.org/CorpusID:5358198} {\bibfield  {journal} {\bibinfo  {journal} {Nano letters}\ }\textbf {\bibinfo {volume} {10 8}},\ \bibinfo {pages} {3129--34} (\bibinfo {year} {2010})}\BibitemShut {NoStop}%
\bibitem [{\citenamefont {Dzsotjan}\ \emph {et~al.}(2010)\citenamefont {Dzsotjan}, \citenamefont {S\o{}rensen},\ and\ \citenamefont {Fleischhauer}}]{PhysRevB.82.075427}%
  \BibitemOpen
  \bibfield  {author} {\bibinfo {author} {\bibfnamefont {David}\ \bibnamefont {Dzsotjan}}, \bibinfo {author} {\bibfnamefont {Anders~S.}\ \bibnamefont {S\o{}rensen}}, \ and\ \bibinfo {author} {\bibfnamefont {Michael}\ \bibnamefont {Fleischhauer}},\ }\bibfield  {title} {\enquote {\bibinfo {title} {Quantum emitters coupled to surface plasmons of a nanowire: A green's function approach},}\ }\href {\doibase 10.1103/PhysRevB.82.075427} {\bibfield  {journal} {\bibinfo  {journal} {Phys. Rev. B}\ }\textbf {\bibinfo {volume} {82}},\ \bibinfo {pages} {075427} (\bibinfo {year} {2010})}\BibitemShut {NoStop}%
\bibitem [{\citenamefont {Ali}\ \emph {et~al.}(2010)\citenamefont {Ali}, \citenamefont {Rau},\ and\ \citenamefont {Alber}}]{PhysRevA.81.042105}%
  \BibitemOpen
  \bibfield  {author} {\bibinfo {author} {\bibfnamefont {M.}~\bibnamefont {Ali}}, \bibinfo {author} {\bibfnamefont {A.~R.~P.}\ \bibnamefont {Rau}}, \ and\ \bibinfo {author} {\bibfnamefont {G.}~\bibnamefont {Alber}},\ }\bibfield  {title} {\enquote {\bibinfo {title} {Quantum discord for two-qubit $x$ states},}\ }\href {\doibase 10.1103/PhysRevA.81.042105} {\bibfield  {journal} {\bibinfo  {journal} {Phys. Rev. A}\ }\textbf {\bibinfo {volume} {81}},\ \bibinfo {pages} {042105} (\bibinfo {year} {2010})}\BibitemShut {NoStop}%
\bibitem [{\citenamefont {Vedral}(2003)}]{PhysRevLett.90.050401}%
  \BibitemOpen
  \bibfield  {author} {\bibinfo {author} {\bibfnamefont {V.}~\bibnamefont {Vedral}},\ }\bibfield  {title} {\enquote {\bibinfo {title} {Classical correlations and entanglement in quantum measurements},}\ }\href {\doibase 10.1103/PhysRevLett.90.050401} {\bibfield  {journal} {\bibinfo  {journal} {Phys. Rev. Lett.}\ }\textbf {\bibinfo {volume} {90}},\ \bibinfo {pages} {050401} (\bibinfo {year} {2003})}\BibitemShut {NoStop}%
\bibitem [{\citenamefont {Wilkens}\ and\ \citenamefont {Meystre}(1991)}]{PhysRevA.43.3832}%
  \BibitemOpen
  \bibfield  {author} {\bibinfo {author} {\bibfnamefont {Martin}\ \bibnamefont {Wilkens}}\ and\ \bibinfo {author} {\bibfnamefont {Pierre}\ \bibnamefont {Meystre}},\ }\bibfield  {title} {\enquote {\bibinfo {title} {Nonlinear atomic homodyne detection: A technique to detect macroscopic superpositions in a micromaser},}\ }\href {\doibase 10.1103/PhysRevA.43.3832} {\bibfield  {journal} {\bibinfo  {journal} {Phys. Rev. A}\ }\textbf {\bibinfo {volume} {43}},\ \bibinfo {pages} {3832--3835} (\bibinfo {year} {1991})}\BibitemShut {NoStop}%
\bibitem [{\citenamefont {Xu}\ \emph {et~al.}(2020)\citenamefont {Xu}, \citenamefont {Xu}, \citenamefont {Theurer}, \citenamefont {Egloff}, \citenamefont {Liu}, \citenamefont {Yu}, \citenamefont {Plenio},\ and\ \citenamefont {Zhang}}]{PhysRevLett.125.060404}%
  \BibitemOpen
  \bibfield  {author} {\bibinfo {author} {\bibfnamefont {Huichao}\ \bibnamefont {Xu}}, \bibinfo {author} {\bibfnamefont {Feixiang}\ \bibnamefont {Xu}}, \bibinfo {author} {\bibfnamefont {Thomas}\ \bibnamefont {Theurer}}, \bibinfo {author} {\bibfnamefont {Dario}\ \bibnamefont {Egloff}}, \bibinfo {author} {\bibfnamefont {Zi-Wen}\ \bibnamefont {Liu}}, \bibinfo {author} {\bibfnamefont {Nengkun}\ \bibnamefont {Yu}}, \bibinfo {author} {\bibfnamefont {Martin~B.}\ \bibnamefont {Plenio}}, \ and\ \bibinfo {author} {\bibfnamefont {Lijian}\ \bibnamefont {Zhang}},\ }\bibfield  {title} {\enquote {\bibinfo {title} {Experimental quantification of coherence of a tunable quantum detector},}\ }\href {\doibase 10.1103/PhysRevLett.125.060404} {\bibfield  {journal} {\bibinfo  {journal} {Phys. Rev. Lett.}\ }\textbf {\bibinfo {volume} {125}},\ \bibinfo {pages} {060404} (\bibinfo {year} {2020})}\BibitemShut {NoStop}%
\bibitem [{\citenamefont {Wiseman}\ and\ \citenamefont {Milburn}(1993{\natexlab{a}})}]{PhysRevLett.70.548}%
  \BibitemOpen
  \bibfield  {author} {\bibinfo {author} {\bibfnamefont {H.~M.}\ \bibnamefont {Wiseman}}\ and\ \bibinfo {author} {\bibfnamefont {G.~J.}\ \bibnamefont {Milburn}},\ }\bibfield  {title} {\enquote {\bibinfo {title} {Quantum theory of optical feedback via homodyne detection},}\ }\href {\doibase 10.1103/PhysRevLett.70.548} {\bibfield  {journal} {\bibinfo  {journal} {Phys. Rev. Lett.}\ }\textbf {\bibinfo {volume} {70}},\ \bibinfo {pages} {548--551} (\bibinfo {year} {1993}{\natexlab{a}})}\BibitemShut {NoStop}%
\bibitem [{\citenamefont {Wiseman}\ and\ \citenamefont {Milburn}(1993{\natexlab{b}})}]{PhysRevA.47.642}%
  \BibitemOpen
  \bibfield  {author} {\bibinfo {author} {\bibfnamefont {H.~M.}\ \bibnamefont {Wiseman}}\ and\ \bibinfo {author} {\bibfnamefont {G.~J.}\ \bibnamefont {Milburn}},\ }\bibfield  {title} {\enquote {\bibinfo {title} {Quantum theory of field-quadrature measurements},}\ }\href {\doibase 10.1103/PhysRevA.47.642} {\bibfield  {journal} {\bibinfo  {journal} {Phys. Rev. A}\ }\textbf {\bibinfo {volume} {47}},\ \bibinfo {pages} {642--662} (\bibinfo {year} {1993}{\natexlab{b}})}\BibitemShut {NoStop}%
\bibitem [{\citenamefont {Li}\ \emph {et~al.}(2008)\citenamefont {Li}, \citenamefont {Zou}, \citenamefont {Shao},\ and\ \citenamefont {Cai}}]{PhysRevA.77.012339}%
  \BibitemOpen
  \bibfield  {author} {\bibinfo {author} {\bibfnamefont {J.~G.}\ \bibnamefont {Li}}, \bibinfo {author} {\bibfnamefont {J.}~\bibnamefont {Zou}}, \bibinfo {author} {\bibfnamefont {B.}~\bibnamefont {Shao}}, \ and\ \bibinfo {author} {\bibfnamefont {J.~F.}\ \bibnamefont {Cai}},\ }\bibfield  {title} {\enquote {\bibinfo {title} {Steady atomic entanglement with different quantum feedbacks},}\ }\href {\doibase 10.1103/PhysRevA.77.012339} {\bibfield  {journal} {\bibinfo  {journal} {Phys. Rev. A}\ }\textbf {\bibinfo {volume} {77}},\ \bibinfo {pages} {012339} (\bibinfo {year} {2008})}\BibitemShut {NoStop}%
\bibitem [{\citenamefont {Carvalho}\ \emph {et~al.}(2008)\citenamefont {Carvalho}, \citenamefont {Reid},\ and\ \citenamefont {Hope}}]{PhysRevA.78.012334}%
  \BibitemOpen
  \bibfield  {author} {\bibinfo {author} {\bibfnamefont {A.~R.~R.}\ \bibnamefont {Carvalho}}, \bibinfo {author} {\bibfnamefont {A.~J.~S.}\ \bibnamefont {Reid}}, \ and\ \bibinfo {author} {\bibfnamefont {J.~J.}\ \bibnamefont {Hope}},\ }\bibfield  {title} {\enquote {\bibinfo {title} {Controlling entanglement by direct quantum feedback},}\ }\href {\doibase 10.1103/PhysRevA.78.012334} {\bibfield  {journal} {\bibinfo  {journal} {Phys. Rev. A}\ }\textbf {\bibinfo {volume} {78}},\ \bibinfo {pages} {012334} (\bibinfo {year} {2008})}\BibitemShut {NoStop}%
\bibitem [{\citenamefont {Steane}(1996)}]{Steane1996TheIT}%
  \BibitemOpen
  \bibfield  {author} {\bibinfo {author} {\bibfnamefont {Andrew~M.}\ \bibnamefont {Steane}},\ }\bibfield  {title} {\enquote {\bibinfo {title} {The ion trap quantum information processor},}\ }\href {https://api.semanticscholar.org/CorpusID:2061791} {\bibfield  {journal} {\bibinfo  {journal} {Applied Physics B}\ }\textbf {\bibinfo {volume} {64}},\ \bibinfo {pages} {623--643} (\bibinfo {year} {1996})}\BibitemShut {NoStop}%
\bibitem [{\citenamefont {Monroe}\ \emph {et~al.}(1995)\citenamefont {Monroe}, \citenamefont {Meekhof}, \citenamefont {King}, \citenamefont {Itano},\ and\ \citenamefont {Wineland}}]{PhysRevLett.75.4714}%
  \BibitemOpen
  \bibfield  {author} {\bibinfo {author} {\bibfnamefont {C.}~\bibnamefont {Monroe}}, \bibinfo {author} {\bibfnamefont {D.~M.}\ \bibnamefont {Meekhof}}, \bibinfo {author} {\bibfnamefont {B.~E.}\ \bibnamefont {King}}, \bibinfo {author} {\bibfnamefont {W.~M.}\ \bibnamefont {Itano}}, \ and\ \bibinfo {author} {\bibfnamefont {D.~J.}\ \bibnamefont {Wineland}},\ }\bibfield  {title} {\enquote {\bibinfo {title} {Demonstration of a fundamental quantum logic gate},}\ }\href {\doibase 10.1103/PhysRevLett.75.4714} {\bibfield  {journal} {\bibinfo  {journal} {Phys. Rev. Lett.}\ }\textbf {\bibinfo {volume} {75}},\ \bibinfo {pages} {4714--4717} (\bibinfo {year} {1995})}\BibitemShut {NoStop}%
\bibitem [{\citenamefont {Cirac}\ and\ \citenamefont {Zoller}(1995)}]{PhysRevLett.74.4091}%
  \BibitemOpen
  \bibfield  {author} {\bibinfo {author} {\bibfnamefont {J.~I.}\ \bibnamefont {Cirac}}\ and\ \bibinfo {author} {\bibfnamefont {P.}~\bibnamefont {Zoller}},\ }\bibfield  {title} {\enquote {\bibinfo {title} {Quantum computations with cold trapped ions},}\ }\href {\doibase 10.1103/PhysRevLett.74.4091} {\bibfield  {journal} {\bibinfo  {journal} {Phys. Rev. Lett.}\ }\textbf {\bibinfo {volume} {74}},\ \bibinfo {pages} {4091--4094} (\bibinfo {year} {1995})}\BibitemShut {NoStop}%
\bibitem [{\citenamefont {Turchette}\ \emph {et~al.}(1995)\citenamefont {Turchette}, \citenamefont {Hood}, \citenamefont {Lange}, \citenamefont {Mabuchi},\ and\ \citenamefont {Kimble}}]{PhysRevLett.75.4710}%
  \BibitemOpen
  \bibfield  {author} {\bibinfo {author} {\bibfnamefont {Q.~A.}\ \bibnamefont {Turchette}}, \bibinfo {author} {\bibfnamefont {C.~J.}\ \bibnamefont {Hood}}, \bibinfo {author} {\bibfnamefont {W.}~\bibnamefont {Lange}}, \bibinfo {author} {\bibfnamefont {H.}~\bibnamefont {Mabuchi}}, \ and\ \bibinfo {author} {\bibfnamefont {H.~J.}\ \bibnamefont {Kimble}},\ }\bibfield  {title} {\enquote {\bibinfo {title} {Measurement of conditional phase shifts for quantum logic},}\ }\href {\doibase 10.1103/PhysRevLett.75.4710} {\bibfield  {journal} {\bibinfo  {journal} {Phys. Rev. Lett.}\ }\textbf {\bibinfo {volume} {75}},\ \bibinfo {pages} {4710--4713} (\bibinfo {year} {1995})}\BibitemShut {NoStop}%
\bibitem [{\citenamefont {Sleator}\ and\ \citenamefont {Weinfurter}(1995)}]{PhysRevLett.74.4087}%
  \BibitemOpen
  \bibfield  {author} {\bibinfo {author} {\bibfnamefont {Tycho}\ \bibnamefont {Sleator}}\ and\ \bibinfo {author} {\bibfnamefont {Harald}\ \bibnamefont {Weinfurter}},\ }\bibfield  {title} {\enquote {\bibinfo {title} {Realizable universal quantum logic gates},}\ }\href {\doibase 10.1103/PhysRevLett.74.4087} {\bibfield  {journal} {\bibinfo  {journal} {Phys. Rev. Lett.}\ }\textbf {\bibinfo {volume} {74}},\ \bibinfo {pages} {4087--4090} (\bibinfo {year} {1995})}\BibitemShut {NoStop}%
\bibitem [{\citenamefont {Barenco}\ \emph {et~al.}(1995)\citenamefont {Barenco}, \citenamefont {Deutsch}, \citenamefont {Ekert},\ and\ \citenamefont {Jozsa}}]{PhysRevLett.74.4083}%
  \BibitemOpen
  \bibfield  {author} {\bibinfo {author} {\bibfnamefont {Adriano}\ \bibnamefont {Barenco}}, \bibinfo {author} {\bibfnamefont {David}\ \bibnamefont {Deutsch}}, \bibinfo {author} {\bibfnamefont {Artur}\ \bibnamefont {Ekert}}, \ and\ \bibinfo {author} {\bibfnamefont {Richard}\ \bibnamefont {Jozsa}},\ }\bibfield  {title} {\enquote {\bibinfo {title} {Conditional quantum dynamics and logic gates},}\ }\href {\doibase 10.1103/PhysRevLett.74.4083} {\bibfield  {journal} {\bibinfo  {journal} {Phys. Rev. Lett.}\ }\textbf {\bibinfo {volume} {74}},\ \bibinfo {pages} {4083--4086} (\bibinfo {year} {1995})}\BibitemShut {NoStop}%
\bibitem [{\citenamefont {Loss}\ and\ \citenamefont {DiVincenzo}(1998)}]{PhysRevA.57.120}%
  \BibitemOpen
  \bibfield  {author} {\bibinfo {author} {\bibfnamefont {Daniel}\ \bibnamefont {Loss}}\ and\ \bibinfo {author} {\bibfnamefont {David~P.}\ \bibnamefont {DiVincenzo}},\ }\bibfield  {title} {\enquote {\bibinfo {title} {Quantum computation with quantum dots},}\ }\href {\doibase 10.1103/PhysRevA.57.120} {\bibfield  {journal} {\bibinfo  {journal} {Phys. Rev. A}\ }\textbf {\bibinfo {volume} {57}},\ \bibinfo {pages} {120--126} (\bibinfo {year} {1998})}\BibitemShut {NoStop}%
\bibitem [{\citenamefont {Knill}\ \emph {et~al.}(2001)\citenamefont {Knill}, \citenamefont {Laflamme},\ and\ \citenamefont {Milburn}}]{Knill2001ASF}%
  \BibitemOpen
  \bibfield  {author} {\bibinfo {author} {\bibfnamefont {Emanuel}\ \bibnamefont {Knill}}, \bibinfo {author} {\bibfnamefont {Raymond}\ \bibnamefont {Laflamme}}, \ and\ \bibinfo {author} {\bibfnamefont {Gerard~J.}\ \bibnamefont {Milburn}},\ }\bibfield  {title} {\enquote {\bibinfo {title} {A scheme for efficient quantum computation with linear optics},}\ }\href {https://api.semanticscholar.org/CorpusID:4362012} {\bibfield  {journal} {\bibinfo  {journal} {Nature}\ }\textbf {\bibinfo {volume} {409}},\ \bibinfo {pages} {46--52} (\bibinfo {year} {2001})}\BibitemShut {NoStop}%
\bibitem [{\citenamefont {Marshall}\ \emph {et~al.}(2009)\citenamefont {Marshall}, \citenamefont {Politi}, \citenamefont {Matthews}, \citenamefont {Dekker}, \citenamefont {Ams}, \citenamefont {Withford},\ and\ \citenamefont {O'Brien}}]{Marshall09}%
  \BibitemOpen
  \bibfield  {author} {\bibinfo {author} {\bibfnamefont {Graham~D.}\ \bibnamefont {Marshall}}, \bibinfo {author} {\bibfnamefont {Alberto}\ \bibnamefont {Politi}}, \bibinfo {author} {\bibfnamefont {Jonathan C.~F.}\ \bibnamefont {Matthews}}, \bibinfo {author} {\bibfnamefont {Peter}\ \bibnamefont {Dekker}}, \bibinfo {author} {\bibfnamefont {Martin}\ \bibnamefont {Ams}}, \bibinfo {author} {\bibfnamefont {Michael~J.}\ \bibnamefont {Withford}}, \ and\ \bibinfo {author} {\bibfnamefont {Jeremy~L.}\ \bibnamefont {O'Brien}},\ }\bibfield  {title} {\enquote {\bibinfo {title} {Laser written waveguide photonic quantum circuits},}\ }\href {\doibase 10.1364/OE.17.012546} {\bibfield  {journal} {\bibinfo  {journal} {Opt. Express}\ }\textbf {\bibinfo {volume} {17}},\ \bibinfo {pages} {12546--12554} (\bibinfo {year} {2009})}\BibitemShut {NoStop}%
\bibitem [{\citenamefont {Lehmberg}(1970{\natexlab{a}})}]{PhysRevA.2.883}%
  \BibitemOpen
  \bibfield  {author} {\bibinfo {author} {\bibfnamefont {R.~H.}\ \bibnamefont {Lehmberg}},\ }\bibfield  {title} {\enquote {\bibinfo {title} {Radiation from an $n$-atom system. i. general formalism},}\ }\href {\doibase 10.1103/PhysRevA.2.883} {\bibfield  {journal} {\bibinfo  {journal} {Phys. Rev. A}\ }\textbf {\bibinfo {volume} {2}},\ \bibinfo {pages} {883--888} (\bibinfo {year} {1970}{\natexlab{a}})}\BibitemShut {NoStop}%
\bibitem [{\citenamefont {Lehmberg}(1970{\natexlab{b}})}]{PhysRevA.2.889}%
  \BibitemOpen
  \bibfield  {author} {\bibinfo {author} {\bibfnamefont {R.~H.}\ \bibnamefont {Lehmberg}},\ }\bibfield  {title} {\enquote {\bibinfo {title} {Radiation from an $n$-atom system. ii. spontaneous emission from a pair of atoms},}\ }\href {\doibase 10.1103/PhysRevA.2.889} {\bibfield  {journal} {\bibinfo  {journal} {Phys. Rev. A}\ }\textbf {\bibinfo {volume} {2}},\ \bibinfo {pages} {889--896} (\bibinfo {year} {1970}{\natexlab{b}})}\BibitemShut {NoStop}%
\bibitem [{\citenamefont {Rudolph}\ \emph {et~al.}(1995)\citenamefont {Rudolph}, \citenamefont {Ficek},\ and\ \citenamefont {Dalton}}]{PhysRevA.52.636}%
  \BibitemOpen
  \bibfield  {author} {\bibinfo {author} {\bibfnamefont {T.~G.}\ \bibnamefont {Rudolph}}, \bibinfo {author} {\bibfnamefont {Z.}~\bibnamefont {Ficek}}, \ and\ \bibinfo {author} {\bibfnamefont {B.~J.}\ \bibnamefont {Dalton}},\ }\bibfield  {title} {\enquote {\bibinfo {title} {Two-atom resonance fluorescence in running- and standing-wave laser fields},}\ }\href {\doibase 10.1103/PhysRevA.52.636} {\bibfield  {journal} {\bibinfo  {journal} {Phys. Rev. A}\ }\textbf {\bibinfo {volume} {52}},\ \bibinfo {pages} {636--656} (\bibinfo {year} {1995})}\BibitemShut {NoStop}%
\bibitem [{\citenamefont {Wang}\ \emph {et~al.}(2005)\citenamefont {Wang}, \citenamefont {Wiseman},\ and\ \citenamefont {Milburn}}]{PhysRevA.71.042309}%
  \BibitemOpen
  \bibfield  {author} {\bibinfo {author} {\bibfnamefont {Jin}\ \bibnamefont {Wang}}, \bibinfo {author} {\bibfnamefont {H.~M.}\ \bibnamefont {Wiseman}}, \ and\ \bibinfo {author} {\bibfnamefont {G.~J.}\ \bibnamefont {Milburn}},\ }\bibfield  {title} {\enquote {\bibinfo {title} {Dynamical creation of entanglement by homodyne-mediated feedback},}\ }\href {\doibase 10.1103/PhysRevA.71.042309} {\bibfield  {journal} {\bibinfo  {journal} {Phys. Rev. A}\ }\textbf {\bibinfo {volume} {71}},\ \bibinfo {pages} {042309} (\bibinfo {year} {2005})}\BibitemShut {NoStop}%
\end{thebibliography}%

\end{document}